\documentclass[12pt]{article}
\usepackage{color,xcolor}
\usepackage[colorlinks,linkcolor=purple,anchorcolor=blue,citecolor=green]{hyperref}
\usepackage{geometry}
\usepackage{wrapfig} 
\usepackage{booktabs}
\usepackage{multirow}
\usepackage{amsmath,amssymb}
\usepackage{amsfonts}
\usepackage{mathrsfs}
\usepackage[T1]{fontenc}

\usepackage{graphicx}
\usepackage{chngcntr}
\counterwithin{figure}{section}
\numberwithin{equation}{section}
\usepackage{subfigure}
\usepackage{amsthm}
\usepackage{enumerate}
\usepackage{bm}
\usepackage{extarrows}
\usepackage{cancel}
\usepackage{lineno}

\usepackage{fancyhdr}
\usepackage{makecell}

\RequirePackage{appendix}

\newtheorem{thm}{Theorem}

\newcommand{\norm}[1]{\|#1\|}

\newcommand{\mbE}{\mathbb{E}}
\newcommand{\mbP}{\mathbb{P}}

\newcommand{\cadlag}{c\`{a}dl\`{a}g }

\title{Piecewise deterministic Markov process for condition-based imperfect maintenance models}
\author{Weikai Wang, Xian Chen\thanks{Corresponding author. Chen's email: chenxian@xmu.edu.cn; Wang's email: weikaiwang@stu.xmu.edu.cn}
	\\ {\small School of Mathematical Sciences, Xiamen University, 361005, P.R. China} \thanks{Funding: This research was supported by the NSFC (grant no. 12171170).}}
\date{}

\begin{document}
	\maketitle
	
	\begin{abstract}
		In this paper, a condition-based imperfect maintenance model based on piecewise deterministic Markov process (PDMP) is constructed. The degradation of the system includes two types: natural degradation and random shocks. The natural degradation is deterministic and can be nonlinear. The damage increment caused by a random shock follows a certain distribution, and its parameters are related to the degradation state. Maintenance methods include corrective maintenance and imperfect maintenance. Imperfect maintenance reduces the degradation degree of the system according to a random proportion. The maintenance action is delayed, and the system will suffer natural degradations and random shocks while waiting for maintenance.
		At each inspection time, the decision-maker needs to make a choice among planning no maintenance, imperfect maintenance and perfect maintenance, so as to minimize the total discounted cost of the system. The impulse optimal control theory of PDMP is used to determine the optimal maintenance strategy. A numerical study dealing with component coating maintenance problem is presented. 
		Relationship with optimal threshold strategy is discussed. Sensitivity analyses on the influences of discount factor, observation interval and maintenance cost to the discounted cost and optimal actions are presented.	
		~
	\end{abstract}

\section{Introduction}\label{section1}

Maintenance is a crucial problem in industrial production. Time-based maintenance (TBM) and condition-based maintenance (CBM) are two categories into which maintenance solutions can be separated based on different standards. Because CBM is more adaptable than TBM and the outcome of condition-based judgment is more accurate than that of time-based judgment, CBM has recently attracted increased attention. Corrective maintenance (CM), also known as replacement or perfect maintenance, is usually carried out when a unit fails. The unit that has performed corrective maintenance will be the same as the new one. However, when it comes to industrial manufacturing, the failure of some components will result in significant losses. Preventive maintenance (PM) is essential to avoiding such system failure, and the preventative maintenance of CBM is currently a popular topic in academic research.
An overview of the optimal maintenance models of condition-based random deterioration systems can be found in \cite{Alaswad1}.

In most CBM models, preventive maintenance is always assumed to be perfect, since this kind of model is much easier to analyze. However, in reality, preventive maintenance is imperfect for a number of reasons, such as the lack of expertise of the maintenance staff, unstable maintenance supplies, and unforeseen events throughout the maintenance process. Therefore, we need to introduce the concept of imperfect maintenance. 
Imperfect maintenance (IM) restores the health condition of a degrading system to any degradation level between as-good-as-new and as-bad-as-old \cite{Pham1}. Generally speaking, The cost of imperfect maintenance will typically be less expensive than the cost of corrective maintenance, however imperfect maintenance will result in more maintenance tasks. Therefore, decision-makers need to make a trade-off between corrective maintenance and imperfect maintenance actions to minimize the total cost of the system. 
The models describing imperfect maintenance are mainly divided into three categories. 
The first type is the minimum preventive maintenance model, in which the system returns to the state it was in prior to the most recent level of degradation after an imperfect maintenance, see \cite{Huynh1}.
The second type is random improvement model, where the reduction of the system's post-maintenance deterioration state is a random variable, see \cite{WuF1}.
The third type is the improvement factor model, in which an improvement factor is multiplied to the system's degradation state after an imperfect maintenance, see \cite{Mosayebi1,LiYanfu1}. 
The improvement factor model was originally introduced by M.A.K. Malik \cite{Malik1}. Since then, numerous scholars have modified it, taking into account things like system longevity, maintenance frequency, inspection time, and other variables. For a review on improvement factor model, see \cite{Mosayebi1}. 
The imperfect maintenance of CBM model has captured the interest of many academics these years. Do et al. \cite{DoP1} build a proactive condition-based maintenance model, considering the perfect and imperfect maintenance strategies. Wang et al. \cite{WangJH1} use the virtual age factor and failure intensity update factor to describe the effect of imperfect maintenance, and build a periodic dynamic imperfect preventive maintenance model applied to the problem of wind turbine maintenance.
Mosayebi Omshi et al. \cite{Mosayebi1} consider the influence of the efficiency of imperfect maintenance on the optimal maintenance strategy. The authors use the improvement factor model to describe the effect of imperfect maintenance, and obtain the optimal threshold strategy by calculating the invariant distribution of the system.

Piecewise deterministic Markov processes (PDMPs) were first introduced by M.H.A. Davis in 1984 \cite{Davis1}. PDMPs form a family of \cadlag Markov processes involving a deterministic motion punctuated by random jumps. In contrast to classical Markov chain models, PDMPs can not only reflect the random jumps of the system, but also depict the deterministic continuous process between jumps. 
In reality, many phenomena need both discrete random jumps and continuous equations to describe. Therefore, PDMPs have been widely applied in various disciplines since it was proposed, such as biology, economics, energy, medicine and other fields, see \cite{ChenXian1,Bauerle1,Defourny1,Dufour1}.

In recent years, PDMPs achieve great attention in maintenance fields, such as \cite{Boukas1,Dufour2,Lair1,Demgne1,LiYanfu2,Arismendi1}. Boukas et al. \cite{Boukas1} consider a maintenance and production model of a flexible manufacturing system using controlled PDMP, and derive the Chapman-Kolmogorov (C-K) equation of the system. But the authors give no numerical result due to the complexity of the model.
Lair et al. \cite{Lair1} build a preventive maintenance model for the maintenance of train air conditioning system using PDMP. They solve the C-K equation of PDMP using the finite volume (FV) method, and give an approximation algorithm of this method. 
Zhang et al. \cite{Dufour2} deal with the offshore oil production using PDMP combined with the simulation method of Monte Carlo.
Demagne et al. \cite{Demgne1} consider a maintenance strategy with stock through PDMP and simulate the system with quasi Monte Carlo method.
Lin et al. \cite{LiYanfu2} develop a maintenance model with cognitive uncertainty and damage interdependence using PDMP. The optimal maintenance strategy is obtained through combined FV method, differential evolution (DE) method and non-dominated sorting differential evolution (NSDE) method. 
Arismendi et al. \cite{Arismendi1} develop a CBM model of critical infrastructure with discrete-state deterioration using PDMP and apply to highway maintenance problems. 
To the best of our knowledge, currently there is \textit{no} existing literature on building \textit{imperfect maintenance} model using PDMP.

In this study, we construct an imperfect maintenance model using PDMP. The degradation state of the unit is continuous. There are two different sorts of degradations: the first is referred to as random shock, which causes discrete random jumps in the degradation state.
The other is called natural degradation, which is deterministic and can be nonlinear. This setting is applicable to systems whose deterioration can be modeled by ordinary differential equations (ODEs) or predicted by statistic methods, see \cite{Fatima1,Arya1,Karanci1} for ODE based modeling and \cite{Panchenko1,Panchenko2} for statistics based modeling. 
In other words, our model is capable of describing a system that simultaneously experiences random shocks and deterministic degradation.

The increment of damage caused by a random shock follows a certain distribution, such as Inverse Gaussian (IG) distribution. The parameters of the IG distribution are related to the jump intensity of the random shocks. The inspection time is fixed. At the time of inspection, the decision-maker will plan the maintenance task according to the current state of the system. The maintenance is not done immediately, but is postponed. The system continues to suffer from random shocks and natural degradation during the waiting time. 

Maintenance methods include corrective maintenance and imperfect maintenance. Corrective maintenance replace the whole unit and restore it into a as-good-as-new status. For imperfect maintenance, we adopt the improvement factor model, which states that imperfect maintenance reduces the degradation degree of the system in a random proportion. The improvement factor obeys the Beta distribution. The efficiency of imperfect maintenance, which is depicted by the parameters of the Beta distribution, is related to the number of executions of imperfect maintenance. The operation of the system and maintenances will incur costs. At each inspection time, the decision-maker needs to make a choice among planning no maintenance, imperfect maintenance and perfect maintenance, so as to minimize the overall discounted cost of the system. 

The major contributions of this paper are as follows: first, the imperfect maintenance model is constructed by using PDMP for the first time; the deterministic natural degradation and random shocks are considered simultaneously.
Second,   the damage increment and the improvement factor depend on the current degradation state of the system and the number of imperfect maintenance respectively,
while most of the existing literature on imperfect maintenance models does not consider such dependences simultaneously, see \cite{WuF1,Mosayebi1,LiYanfu1,WangJH1}. 
Finally, a type of impulse control theory in \cite{Costa1} is applied since the existing methods used to deal with PDMP in maintenance models are not suitable for our work. In particular, the widely used C-K equation is not appropriate for our model because the C-K equation derived would be very complicated and difficult to solve.

The rest of the article is arranged as follows: Section \ref{section3} gives the imperfect maintenance model based on PDMP, optimal impulse control theory and related calculations. In Section \ref{section4}, PDMP model is applied to a coating maintenance problem as an illustrative example; numerical simulation is performed; a discussion on our method and the threshold method is presented. Section \ref{section5} analyzes the sensitivity of some parameters. Section \ref{section6} summarizes the content of this paper and prospects future research.

\section{Piecewise deterministic Markov processes}\label{section3}
\emph{Notations}. In this article, $\mathbb{R}$ denotes the set of all real numbers. $\mathbb{R}_+ = [0,+\infty)$ is the set of all non-negative real numbers. $\mathbb{N}=\{ 0,1,2,\cdots \}$ is the set of all natural numbers. $\delta_x(A)$ is the Dirac measure of a set $A\subset X$, which equals $1$ if $x\in A$ and $0$ otherwise. $1_A$ is the indicator function of a set $A$. 
$\lfloor \cdot \rfloor$ is the rounding down operator, $\mbE(\cdot)$ is the expectation and $\mathrm{Var}(\cdot)$ is the variance.

In this section, we describe how to build an imperfect maintenance model based on PDMP. We first give the fundamental assumptions of the model, then introduce the degradation-maintenance model, which includes the degradation process, maintenance process, inspection process and waiting process. After that, we give a serious mathematical construction of piecewise deterministic Markov decision process (PDMDP) and the optimal impulse control theory. Finally, we provide the computation of the value function.

\subsection{Mathematical background}
A PDMP can be described using the state space $X$, the action space $A$, the flow $\phi$, the jump intensity $\eta$ and the transition measure $Q$ . We define $X\subset\mathbb{R}^k$ , $k\in\mathbb{N}$, as the state space and $\partial X$ as the boundary. The flow is defined as $\phi(x,t):\mathbb{R}^k\times\mathbb{R}\mapsto \mathbb{R}^k$, it describes the deterministic trajectory of the system between jumps. The active boundary is defined as $\Xi=\{ x\in\partial X: x=\phi(y,t),y\in X,t\in\mathbb{R}_+ \}$. We denote the extended state space as $\bar{X} = X\cup\Xi$ and the time for the flow to hit the boundary starting from  $x$ as $t^*(x) = \inf\{ t\in\mathbb{R}_+:\phi(x,t) \in \Xi \}$. $A$ is the action space. Since the system would take different actions at flow and boundaries, we define the set of state-action pairs as $\mathbb{K} = \mathbb{K}^g\cup\mathbb{K}^i$, where $\mathbb{K}^g$ and $\mathbb{K}^i$ are defined respectively as
\begin{align*}
\mathbb{K}^g &= \{ (x,a)\in X\times A: a\in A^g(x)\},\\
\mathbb{K}^i &= \{ (x,a)\in \Xi\times A: a\in A^i(x)\}.
\end{align*}
Here $A^g(x)$ and $A^i(x)$ are the sets of available actions when the system is at flow and boundaries respectively. The jump intensity $\eta:\mathbb{K}^g \mapsto\mathbb{R}_+$ is a measurable function. $q$ is a signed kernel on $X$ given  $\mathbb{K}^g$. It represents the transition rate of random jumps and satisfies $q(X|x,a) = 0$. Let $\eta(x,a) = -q(\{x\}|x,a) = q(X\backslash\{x\}|x,a)$. $Q$ is a random kernel on $X$ given $\mathbb{K}^i$. It determines the distribution of states of the system after jumps at boundaries. Suppose the system starts at a pre-jump state $x$, and after time $t^*(x)$ it hits the boundary. The current state of the system is $\phi(x,t^*(x))$. The decision-maker can make a decision by choosing an action $a$ from the action set $A(\phi(x,t^*(x)))$. The system then jumps to a new state $x'$ through transition measure $Q(\cdot|\phi(x,t^*(x)),a)$. Note that the distribution of states after random jump can be defined as $Q(dy\backslash\{x\}|x,a) = q(dy\backslash\{x\}|x,a)/\eta(x,a)$, $x\in X$. So the transition rate can be defined using $Q$ as 
\[ q(dy|x,a) = \eta(x,a) [ Q(dy|x,a) - \delta_x(dy)], \  \forall x\in \bar{X}.   \]

\subsection{Model assumptions}
Our model is mainly based on the following assumptions:
\begin{itemize}
	\item The deterioration of the unit can be monitored with numerical value, and the observed information reveals the true degradation state of the unit.
	
	\item The inspection is periodic. The decision-maker gets full information about the unit at the inspection time and the maintenance time. The \textit{observation} refers to the regular inspection and the maintenance. At any other time, although the state of the system is changing, the decision-maker cannot know the state of the system.
	
	\item The deterioration and failure of the unit cannot be fixed by itself, and the degradation state of the unit can only be known by observations, that is, if the unit is failed between two observations, it will remain in the failure state until the next observation.
	
	\item There is a time elapsed from the time of planning a maintenance action to the time of executing the maintenance action, which means the maintenance action is delayed. In fact, due to the need of preparing devices and equipments, such as ordering corresponding parts, the maintenance action is usually postponed. In order to simplify the calculations, we assume that the time of maintenance delay is fixed. 
	
	\item The maintenance duration is neglected in this model.
\end{itemize}

\subsection{Degradation-maintenance model}
We use a five tuple $x=(w,n,\sigma,d,\theta)$ to represent the state of the system. The continuous real variable $w\in\mathbb{R}_+$ denotes the degradation of the unit. $w=0$ represents the state of no degradation, $w \ge M$ represent the state of failure. $n\in\mathbb{N}$ is the number of executed imperfect maintenance after last corrective maintenance (replacement). $\sigma$ is the time since the last inspection or maintenance until now. Discrete integer variable $d$ represents the planned action. $\theta$ is the time from the beginning till present. We denote the state space of the system as $X = \{ x: x=(w,n,\sigma,d,\theta)\in \mathbb{R}_+\times\mathbb{N}\times\mathbb{R}_+\times\mathbb{N}\times \mathbb{R}_+ \}$.

We introduce the construction of degradation-maintenance model in detail in the following subsections. 

\subsubsection{Degradation process}
There are two sources of degradation to the unit: one is the natural degradation caused by the increase of operating time, and the other is the random shocks caused by accidents. We first introduce the natural degradation, and then describe the random shocks.

When the unit is functioning normally, wear is caused by its operation, and the degree of damage steadily increases until the unit fails. This type of damage is referred to as natural degradation. The unit's natural degeneration may typically be predicted in advance. For instance, the manufacturer will test the product's operational lifespan and durability, and based on the test findings, the wear of the product under normal use can be determined. Alternatively, the unit's natural degradation can be determined directly using physical formulas, so that the natural deterioration in various stages can be estimated.

In our model, we assume that the natural degradation can be calculated using $\varphi(x,t)$. If the starting state of the system is $x=(w,n,\sigma,d,\theta)$, under the situation of no random shocks and external interferences, after time $t$ the degradation state of the unit is $x' = \varphi(x,t)$. Generally speaking, the natural degradation rate of a unit is related to the degradation state of the unit. The rate of natural degradation increases with a unit's degree of degradation. We suppose $\varphi(x,t)$ is monotonically increasing in $w$. When the degradation state exceeds $M$, the unit is in the failed state. We let $w=M$, and until the next maintenance, the degradation state of the unit remains unchanged.

Random shocks will make the unit enter a more serious degradation state. We suppose the jump intensity of  random shocks is $\eta(x)$. The intensity is related to the current state of the system $x$. In our model, $\eta$ is not related to action $a$. We assume that the more serious the degradation state of a unit is, the greater the probability and the higher the frequency of a random shock occurs. From a mathematical point of view, $\eta:X\to\mathbb{R}^+$ is a measurable function, and satisfies the following properties: for any $x\in X$, there exists $\varepsilon > 0$ such that
\[ \int_0^\varepsilon \eta(\phi(x,t)) dt := \int_0^\varepsilon \eta(\varphi(x,t),n,\sigma+t,d,\theta+t)dt < \infty, \]
where $\phi(x,t)$ is the state of the system after time $t$ starting from $x$. Let $\Lambda(x,t) = \int_0^t \eta(\phi(x,s))ds$, $t < t^*(x)$. Here $t^*(x)$ is the time for the system starting from $x$ to reaching the boundary of the flow. Then, starting from state $x$, the first jump time $T_1$ is determined by the following survival function:
\[ \mbP_x(T_1 > t) = e^{-\Lambda(x,t)} 1_{\{ t<t^*(x) \}}.  \]

When the shock happens, the degradation state of the unit will increase by a random variable. Inverse Gaussian process is generally used to represent the random damage of the unit, see \cite{YeZS1,Mosayebi1}. In our model, we assume that the damage increment of the unit at the time of a random shock follows the IG distribution. Its probability density function (pdf) is given by:
\begin{equation}\label{eq-IG-pdf}
f_{IG}(\varpi,h) = \sqrt{\frac{\lambda h^2}{2\pi \varpi^3}}\exp\left\{ -\frac{\lambda (\varpi-\mu h)^2}{2\mu^2 \varpi} \right\},\quad \varpi>0,
\end{equation}
where $\mu h$ is the mean and $\lambda h^2$ is the shape parameter of the IG distribution. The constants $\mu > 0$, $\lambda > 0$ are determined in advance. $h = 1/\eta(x^-)$ is the average time interval of random jumps when the shock occurs where $x^-$ is the state just before the shock occurs.

Of course, the random damage can also be set as other random variables, such as variables following Gamma distribution, geometric distribution, see \cite{WuD1,Alaswad1}.

\subsubsection{Maintenance process}
In our model, we consider two kinds of maintenance actions: imperfect maintenance and corrective maintenance (replacement or perfect maintenance).

We use the random improvement factor model to describe the imperfect maintenance process. According to \cite{Mosayebi1}, the degradation state after performing imperfect maintenance at time $t$ can be given by:
\[ w_t = \vartheta w_{t-}.  \]
where  $0<\vartheta < 1$ is the improvement factor, and $w_{t-}$  is the degradation state before the maintenance action. $\vartheta$ is a random variable,  which indicates the effect of imperfect maintenance. In our model, we assume that $\vartheta$ follows Beta distribution with parameters $\alpha(n_{t-}),\beta$, where $\alpha$ is monotonically increasing in $n_{t-}$. $n_{t-}$ is the number of executed imperfect maintenance after last corrective maintenance. The pdf of $\vartheta$ is given by
\begin{equation}\label{eq-Beta-pdf}
f_{\mathrm{B}}(\vartheta,n_{t-}) = \frac{\vartheta^{\alpha(n_{t-})-1}(1-\vartheta)^{\beta-1}}{B(\alpha(n_{t-}),\beta)},\quad B(\alpha(n_{t-}),\beta) = \frac{\Gamma(\alpha(n_{t-}))\Gamma(\beta)}{\Gamma(\alpha(n_{t-})+\beta)}.
\end{equation}
The reason why Beta distribution is used is that Beta distribution is often used to describe the effect of imperfect maintenance, see \cite{Mosayebi1,LiYanfu1}. It can describe many types of density functions through only two parameters, such as increasing, decreasing, U-shaped, and inverse U-shaped. The expectation and variance of $\vartheta$ are $\mbE(\vartheta) = \frac{\alpha(n_{t-})}{\alpha (n_{t-})+\beta}$, $\mathrm{Var}(\vartheta) = \frac{\alpha(n_{t-}) \beta}{(\alpha(n_{t-})+\beta)^2(\alpha(n_{t-}) +\beta+1)}$. $\mbE(\vartheta)$ can indicate the efficiency of imperfect maintenance. It is easy to see that as $n_{t-}$ increases, $\mbE(\vartheta)$ gradually approaches $1$, the efficiency of imperfect maintenance will become lower. This means that the more imperfect maintenance is performed, the worse the maintenance effect is. The pdf and cdf shapes of Beta distribution under different parameters and the influence of parameters of Beta distribution on maintenance efficiency can be referred to \cite{Mosayebi1}. The variance $\mathrm{Var}(\vartheta)$ can characterize the stability of imperfect maintenance results.

Corrective maintenance is also called perfect maintenance or replacement, that is, the unit is in a as-good-as-new state after maintenance, and is equivalent to replacing it with a new unit. Compared with imperfect maintenance, the cost of corrective maintenance is usually higher.

\subsubsection{Inspection and waiting process}

We suppose that the system has a fixed inspection time $T_{\mathrm{isp}}$. The decision-maker fully recognizes the state of the unit at inspection time, and makes corresponding decisions according to the state: no maintenance, imperfect maintenance or corrective maintenance. If the decision-maker plans to maintenance, then after a waiting time $T_{\mathrm{soj}}$, the maintenance action is performed. We assume that $T_{\mathrm{soj}} < T_{\mathrm{isp}}$, which means that the waiting time for maintenance should be less than the inspection interval. We denote $\sigma$ as the time since the last observation (inspection or maintenance). At every inspection time, let $\sigma = 0$. After any maintenance action, we suppose the decision-maker fully knows the state after the maintenance. So it is equivalent to performing an observation, we let $\sigma = 0$ and count from the current time. If the decision-maker does not plan any maintenance action, the system will continue to operate until the next inspection time.

Since there is a delay between making a choice and implementing a maintenance plan, the unit's degradation status will continue to worsen during this time. As a result, the decision-maker must take into account both the condition of the unit at the moment of inspection and the condition after some waiting time. If the planned maintenance action is imperfect maintenance, and the unit is in failed state when the maintenance task is to be performed after the waiting time, the maintenance action is changed to corrective maintenance.

\begin{figure}[htbp]
	\centering
	\includegraphics[scale=0.6]{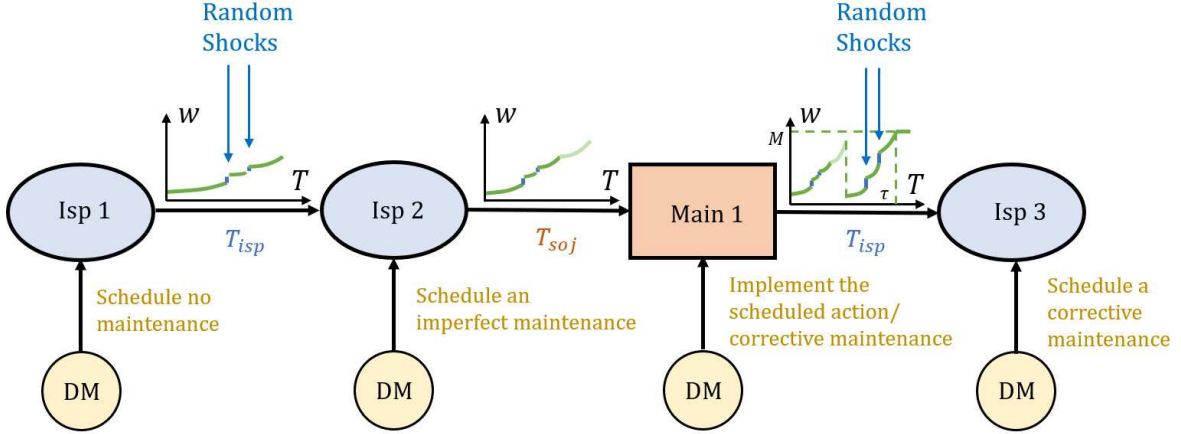}
	\caption{\it An illustration of the degradation-maintenance process. }
	\label{PDMP-1}
\end{figure}

Figure \ref{PDMP-1} provides an illustration of the degradation-maintenance process. The unit deteriorates as a result of random shocks and natural deterioration. The green curves represent the unit's state of degradation. The blue dot lines stand for random jumps. $w$ represents the degradation state and $T$ stands for the time in the coordinate system.
The decision-maker (DM) at Isp 1 has no maintenance actions planned. The DM schedules an imperfect maintenance at Isp 2. The scheduled action is carried out at Main 1. If the unit breaks down at Main 1, the DM should put corrective maintenance into action. The system breaks down at some time $\tau$ before Isp 3 and remains in the broken state $M$. The DM schedules a corrective maintenance at Isp 3.

\subsection{Formulation of the piecewise deterministic Markov decision process model}

After constructing the degradation-maintenance model, we need to formulate the piecewise deterministic Markov decision process (PDMDP) model. In the above model, the deterministic flow $\phi$ between jumps, the spontaneous jump caused by random shocks of the system and the impulse jumps through maintenance process are defined. We still need to define the boundary set $\Xi$, the action set $A$ and the transition measure $Q$. In this subsection, the rigorous definitions of the parameters of PDMDP are presented.

\subsubsection{Actions}
The actions that the decision-maker can choose in the whole system are: no maintenance, imperfect maintenance and perfect maintenance, which are denoted by $0,1,2$ respectively. The action set is defined as $A = \{ 0,1,2\}$.

\subsubsection{Boundaries}
Boundaries are crucial to impulse control of PDMP. Based on the above model, we have the following five boundaries of the system:
\begin{itemize}
	\item When the simulation ends, we need to form a boundary $\Xi_{\mathrm{end}} = \{ x: \theta = T_{\mathrm{end}} \}$ where $T_{\mathrm{end}}$ is the end time of the simulation. At this boundary, the decision-maker does not need to make any decision.
	
	\item When the system reaches its inspection time, a boundary is given by $\Xi_1 = \{ x: \sigma = T_{\mathrm{isp}} \}$. At this boundary, the decision-maker needs to make a decision. The action set is
	$A(x) = \{ 0,1,2\}$, $x\in\Xi_1$.
	
	\item When the decision-maker scheduled a maintenance action, and after a waiting time, a new boundary forms. Since we need to consider different situations during maintenance, we form three boundaries:
	$\Xi_{21}=\{ x: \sigma = T_{\mathrm{soj}}, d=1,w<M\}$ denoting the boundary that the scheduled action is imperfect maintenance and at the maintenance time, the unit is operating normally; 
	$\Xi_{22}=\{x: \sigma = T_{\mathrm{soj}}, d=2\}$ denoting the boundary that the scheduled action is corrective maintenance; $\Xi_{23}=\{ x:\sigma = T_{\mathrm{soj}}, d=1,w\ge M \}$ denoting the boundary that the scheduled action is imperfect maintenance and at maintenance time the system is in the failure state. At these three boundaries, the action set is $A(x) = \{0\}$, $x\in\Xi_{21}\cup\Xi_{22}\cup\Xi_{23}$.
\end{itemize}

Set $\Xi = \Xi_{\mathrm{end}}\cup\Xi_1\cup\Xi_{21}\cup\Xi_{22}\cup\Xi_{23}$. We summarize the actions that the decision-maker can take in all states as
\begin{equation*}
A(x) = \begin{cases}
\{ 0 \}, & x\in \bar{X}\backslash \Xi_1,\\
\{0,1,2\}, & x\in \Xi_1.\\
\end{cases}
\end{equation*}

\subsubsection{Deterministic process}
Between two jumps, the system evolves through the flow $\phi(x,t)$. When the starting state is $x\in X$, $t < t^*(x)$,
\[ \phi(x,t) = (\varphi(x,t),n,\sigma + t,d,\theta + t).\]
The formula of $\varphi$ is predetermined, and is monotonically increasing in $t$. When $w\ge M$, the unit is in the failure state, let $\varphi(x,t)=M$ until the flow hits the next boundary. Let $\Delta$ be an absorbing state, denoting the end of the simulation. For any $t\in\mathbb{R}_+$, we have $\phi(\Delta,t) = \Delta$.

\subsubsection{Random jumps}
Random shocks will cause random jumps in the system. Assume that the jump intensity of the random shock is $\eta(x)$. When the shock happens, the increment of damage of the unit $\varpi\in \mathbb{R}_+$ subjects to IG distribution with parameter $\mu h$ and $\lambda h^2$. Suppose the shock happens at time $t$, the state of the system  before the shock is denoted as $x_{t-} = (w_{t-},n_{t-},\sigma_{t-},d_{t-},\theta_{t-})$. When the shock happens, a realization of  the increment  of damage $\varpi$ is denoted as $\hat{\varpi}$, the degradation state of the system increases, $w_t = w_{t-} + \hat{\varpi}$, other parameters $n_t,\sigma_t,d_t,\theta_t$ are the same as before.

\subsubsection{Jumps at intervention times and maintenance costs}
There will be deterministic jumps during inspections and maintenances. At some inspection time $t$, the state of the system before inspection is denoted as $x_{t-} = (w_{t-},n_{t-},\sigma_{t-},d_{t-},\theta_{t-})$. At this time, the decision-maker makes decisions according to the state of the system, chooses an action $a$ from the action set $A(x_{t-})$. The variable $d$ records the action made by the decision-maker, which means $d_t=a$. Meanwhile, let the inspection counting variable $\sigma_t = 0$.

If $d_t = 0$, it means that the planned action is no maintenance. The system continues and the decision-maker makes decisions at the next inspection time $t' = t + T_{\mathrm{isp}}$. If $d_t = 1,2$, it means that the planned action is maintenance. The system continues and the maintenance action is executed when $t' = t+T_{\mathrm{soj}}$. The state after maintenance is $x_{t'} = (w',n',\sigma',d',\theta')$, where $\sigma' = 0$, $d' = 0$, $\theta' = \theta_t + T_{\mathrm{soj}}$.

If $d_t=2$, it means that the planned action is corrective maintenance. The degradation state and number of executed imperfect maintenance change to $w' = 0$, $n' = 0$. The cost of corrective maintenance is a constant $C_2$.

If $d_t=1$, it means that the planned action is imperfect maintenance. The degradation state and number of executed imperfect maintenance change to $w' = \hat{\vartheta}w_{t'-}$, $n' = n_t + 1$, where $\hat{\vartheta}$ is a realization of $\vartheta$. Costs will be incurred while performing maintenance actions. Denote the cost of imperfect maintenance as $C_1$. It consists of a fixed cost $c_1$ and a changing cost $\hat{c}_1$. That is $C_1(x_{t'}) = c_1 + \hat{c}_1(x_{t'-})$, where $\hat{c}_1(x_{t'-})$ is related to the degradation state and number of executed imperfect maintenances before maintenance. We suppose that $\hat{c}_1$ is monotonically increasing with $w_{t'-},n_{t'-}$.

If the system is in the failure state at the time of maintenance, which means $w_{t'-} \ge M$, but the planned action is imperfect maintenance, the executed action is changed to corrective maintenance. In reality, temporary changes of the maintenance plan will result in additional costs, such as urgent orders, requiring additional maintenance personnel, etc. The maintenance cost incurred in this situation is denoted as $C_3$, which is higher than the cost of corrective maintenance $C_2$. In general, we have $C_1 \le C_2 < C_3$. Although the cost of imperfect maintenance is less expensive than corrective maintenance, the uncertainty of the result of imperfect maintenance will lead to the increase of maintenance frequency. To reduce the overall maintenance cost, the decision-maker need to trade off between repeated imperfect maintenance and one-time corrective maintenance.

\subsubsection{Transition measure}
Based on the maintenance action and the deterministic process of the system, we can give the transition measure of the system at boundaries as follows. Assume that the state of the system before hitting the boundary is $x=(w,n,\sigma,d,\theta)$, we have
\begin{equation}\label{eq-impulse-jump}
\begin{aligned}
Q(dy|x,a) =&1_{\{x\in\Xi_{\mathrm{end}}\}}\delta_{(w,n,\sigma,d,\theta)}(dy)+1_{\{x\in\Xi_{1}\}}\delta_{(w,n,0,a,\theta)}(dy)\\
&+1_{\{ x\in\Xi_{21}\}} \int_0^1 \delta_{(\vartheta w, n+1,0,0,\theta)}(dy) f_{\mathrm{B}}(\vartheta,n)d\vartheta\\
&+1_{\{ x\in\Xi_{22}\}} \delta_{(0,0,0,0,\theta)}(dy) + 1_{\{ x\in\Xi_{23}\}} \delta_{(0,0,0,0,\theta)}(dy).
\end{aligned}
\end{equation}
where $f_\mathrm{B}$ is defined by (\ref{eq-Beta-pdf}). Based on the degradation process, for any $x\in\bar{X}$, the transition rate of the spontaneous random jumps of the system is given by
\begin{equation}\label{eq-natural-jump}
q(dy|x,a) = \eta(x) \left[ \int_0^\infty \delta_{(w + \varpi,n,\sigma,d,\theta)} (dy) f_{\mathrm{IG}}(\varpi,1/\eta(x)) d\varpi - \delta_{\{x\}}(dy) \right].
\end{equation}
where $f_{\mathrm{IG}}$ is defined as (\ref{eq-IG-pdf}).

\subsubsection{Cost function}
The cost function of the system has two parts: one part is the impulse control cost $C^i$, and the other part is the continuous operating cost $C^g$. The impulse control cost, which is the maintenance cost plus the inspection cost mentioned above, is defined as
\[ C^i(x) = 1_{\{x\in\Xi_{21}\}}C_1(x) + 1_{\{x\in\Xi_{22}\}} C_2 + 1_{\{ x\in \Xi_{23}\}} C_3 + 1_{\{x\in\Xi_1\}}C_{\mathrm{isp}}. \]
where $C_{\mathrm{isp}}$ is the fixed inspection cost.

The long-term operation of the system at a high level of degradation will lead to many risks, especially for systems with high safety requirements, such as nuclear power plant systems, one failure will lead to irreparable consequences. Thus, we determine a threshold $\zeta$. When the degradation level of the unit is higher than the threshold $\zeta$, additional continuous operating costs will be incurred, defined as
\[ C^g(x) = C_g + 1_{\{ w > \zeta \}}C_{\mathrm{grd}}(w-\zeta),\]
where $C_g$ is the fixed operating cost,  and $C_{\mathrm{grd}}$ is the additional operating cost. We assume that $C_{\mathrm{grd}}$ is monotonically increasing in $w$.

The strategy $u$ is a sequence of functions $u = \{u_k\}_{k\in\mathbb{N}}$, where $k$ denotes the $k$th decision made by the decision-maker. $u_k: \Xi_1\mapsto A$, it specifies what actions the decision-maker will take at the moment of  $k$th  decision according to the current state $x\in \Xi_1$. For a precise definition, please refer to \cite{Costa1} Section 2.3. Define the set of all possible strategies as $\mathcal{U}$. Suppose the system starts at state $x_0$, the state of the system at time $t$ is denoted by $x_t$. Then for any available strategy $u\in\mathcal{U}$, based on the continuous operating cost and impulse control cost above, we can define the discounted cost value function of the system as
\[ V(u,x) = \mbE_x^u \left[ \int_{(0,\infty)} e^{-\rho s} C^g(s)ds \right] + \mbE_x^{u}\left[ \int_{(0,\infty)} e^{-\rho s} 1_{\{x_{s-}\in\Xi\}} \int_{A(x_{s-})} C^i(x_{s-},a) u(da|x_{s})\nu(d{s}) \right]. \]
where $x_{s-}$ is the pre-jump state at time $s$. $\rho>0$ is the discount factor and $\nu$ is a measure that counts the number of jumps in the process.

\subsection{Optimal impulse control}
The optimal impulse control theory of PDMP was first given by Costa and Davis \cite{Costa2} in 1989. In recent years, new studies on impulse control of PDMP have been conducted, which can cope with border types of controls and have more effective computational algorithms, see \cite{Dufour3,Costa1,Dufour4}. 

The following theorem is from \cite{Costa1} Theorem 5.5. It can be used to determine the optimal cost and optimal control strategy of the system through iterative equations.
\begin{thm}
	\label{thm:1}
	Suppose Assumptions A,B,C in \cite{Costa1} Section 3.2 are satisfied. For any $x\in\bar{X}$ we define the sequence of functions  $\{ W_k \}_{k\in\mathbb{N}}$ as follows: for any $x\in\bar{X}$ 
	\begin{equation}\label{thm1-1}
	\begin{aligned}
	W_0(x) &= K_A 1_{A_{\varepsilon_1}}(x) + (K_A + K_B) 1_{A_{\varepsilon_1}^c} (x),\\
	W_{k+1}(x) &= \mathfrak{B} W_k(x),\quad k\in \mathbb{N},
	\end{aligned}
	\end{equation}
	where $K_A$ and $K_B$ are constants defined in \cite{Costa1} Lemma 5.3, $A_{\varepsilon_1} = \{ x\in X: t^*(x) > \varepsilon_1 \}$ and $K\ge \sup_{x\in X}\eta(x)$ is a constant.
	\[ \mathfrak{B} V(y) = \int_{[0,t^*(y)]} e^{-(K+\rho)t} \mathfrak{R} V(\phi(y,t)) dt + e^{-(K+\rho) t^*(y)} \mathfrak{T} V(\phi(y,t^*(y))),  \]
	where $\rho>0$ is the discount factor. For any $V$ defined on $\bar{X}$, $\mathfrak{R}V$ and $\mathfrak{T}V$ are defined as follows: 
	\begin{equation}\label{thm1-2}
	\begin{aligned}
	\mathfrak{R}V(x) &= C^g(x) + q V(x) + \eta V(x),\quad x\in X,\\
	\mathfrak{T}V(z) &= \inf_{a\in A(z)} \left\{ C^i(z,a) + QV(z,a) \right\},\quad z\in\Xi.
	\end{aligned}
	\end{equation}
	where $q$ is the transition rate of the random shocks, $Q$ is the probability distribution after the impulsed jumps, defined as (\ref{eq-natural-jump}) and (\ref{eq-impulse-jump}) respectively. The sequence of functions $\{W_k\}_{k\in\mathbb{N}}$ converges to a function $W$ defined on $\bar{X}$.  Moreover, we have
	\begin{enumerate}[(i)]
		\item $W(x) = \inf_{u\in \mathcal{U}} V(u,x)$, $\forall x\in \bar{X}$;
		\item There is a measurable mapping $\hat{\varphi}:\Xi\to A$ such that $\hat\varphi(z)\in A(z)$ for any $z\in \Xi$ and satisfying
		\[ C^i(z,\hat{\varphi}(z)) + QW(z,\hat{\varphi}(z)) = \inf_{a\in A(z)}\left\{ C^i(z,a) + QW(z,a) \right\}. \]
	\end{enumerate}
\end{thm}

Roughly speaking, in the above theorem, $\mathfrak{R}$  and $\mathfrak{T}$ describe  one step change of value function of gradual cost and impluse  cost respectively.  And $\mathfrak{B}$ is the combination of $\mathfrak{R}$  and $\mathfrak{T}$, such that the fixed point of the equation W=$\mathfrak{B}$W is the optimal value function, hence 
iterative sequence $W_k=\mathfrak{B}^kW_0$ will converge to the optimal value function. Interested readers could refer to \cite{Costa1} for more information.

Based on Theorem \ref{thm:1}, starting from any state $x_0$, equation (\ref{thm1-1}) finally converges to a function $W$ and its value equals $\inf_{u\in \mathcal{U}}V(u,x_0)$. Let $\varepsilon>0$ be a given stopping limit. In computation, when $\norm{W_{k+1}(x) - W_k(x)}_\infty < \varepsilon$, we can assume that (\ref{thm1-1}) converges. Thus for the above $W_{k+1}$, for any given state $x\in\Xi$, we can calculate $\arg\min_{a\in A(x)} \{ C^i(x,a) + QW_{k+1}(x,a)\}$ to get the approximate optimal actions.

In the following subsection, we give the computation of the above $\mathfrak{B}V$ in our model.

\subsection{Computation of the value function}
\subsubsection{Computation of $\mathfrak{R}$}
For any $x=(w,n,\sigma,d,\theta)\in X$ and $V:\bar{X}\mapsto \mathbb{R}$, $\mathfrak{R}V(x)$ is defined as:
\[ \mathfrak{R}V(x) = C^g(x) + qV(x) + \eta V(x), \]
where $q$ computes the difference between the states before and after the spontaneous jump. According to the above model, we have
\[ q(dy|x,a) = \eta(x) \left[ \int_0^\infty \delta_{(w + \varpi,n,\sigma,d,\theta)} (dy) f_{\mathrm{IG}}(\varpi,1/\eta(x)) d\varpi - \delta_{\{x\}}(dy) \right], \]
where $\varpi > 0$. For the convenience of computation, we divide the possible random damage increment into finite intervals $\{U_1,U_2,\cdots,U_k\}$, $U_s=[u_s,u_{s+1})$, $s=1,\cdots,k-1$, where $U_k=[u_k,\infty)$, $k\in\mathbb{N}$. Take a value in the interval to represent the random damage increment value of this interval, and its probability is the cumulative probability distribution of this interval. For instance, $\varpi_s = \frac{1}{2}(u_s+u_{s+1}) \in U_s$, $s=1,\cdots,k-1$, $\varpi_k = u_k$. Its probability is defined as
\begin{align*}
\hat{f}_{\mathrm{IG}}(\varpi_s,1/\eta(x)) := \int_{U_s} f_{\mathrm{IG}}(\varpi,1/\eta(x)) d\varpi,\quad 1\le s \le k.
\end{align*}
For notational convenience, we write above $\varpi_s\in U_s$ as $\varpi\in\varPi$. $\varPi$ is the finite set of all $\varpi_s$, which includes all possible values of damage increment and can be determined in advance by experiments. We assume that $0 \notin \varPi$. Thus the above $qV(x)$ can be written as
\[ qV(x) = \int_X V(y) q(dy|x) \approx \eta(x) \left[ \sum_{\varpi\in\varPi} V(w+\varpi,n,\sigma,d,\theta) \hat{f}_{\mathrm{IG}}(\varpi,1/\eta(x))  - V(x) \right].  \]
Then we have the expression of $\mathfrak{R}V(x)$:
\begin{align*}
\mathfrak{R}V(x) &= C^g(x) + qV(x) + \eta V(x)\\
&\approx C_g + 1_{\{ w > \zeta \}} C_{\mathrm{grd}}(w-\zeta) + \eta(x) \sum_{\varpi\in\varPi} V(w+\varpi,n,\sigma,d,\theta) \hat{f}_{\mathrm{IG}}(\varpi,1/\eta(x)),\\
\mathfrak{R}V(\Delta) &= 0.
\end{align*}

\subsubsection{Computation of $\mathfrak{T}$}
Similar to the computation of $\mathfrak{R}$, we calculate
\begin{align*}
\mathfrak{T}V(x) = \inf_{a\in A(x)} \left\{ C^i(x,a) + QV(x,a) \right\},  \forall x\in\Xi.
\end{align*}
According to the transition measure, similar to the definition of $\varpi$, we can divide the interval $[0,1]$ into finite intervals $\{V_1,\cdots,V_k\}$. Take the median value of each interval to represent the imperfect maintenance factor of that interval. Then for $\vartheta_s \in V_s$, its probability is the cumulative probability distribution of $\vartheta$ in that interval:
\[ \hat{f}_B(\vartheta_s,n) := \int_{V_s} f_B(\vartheta,n) d\vartheta,\quad 1\le s\le k. \]
For notational convenience, we can write $\vartheta_s\in V_s$ as $\vartheta \in \varTheta$. $\varTheta$ is the finite set of all possible imperfect maintenance factors and can be determined in advance by experiments. Thus the expression of $QV(x,a)$ can be written as:
\begin{align*}
QV(x,a) =& \int_X V(y) Q(dy|x,a) \approx 1_{\{ x\in\Xi_{\mathrm{end}} \}} V(w,n,\sigma,d,\theta)\\
&+ 1_{\{x\in\Xi_1 \}} V(w,n,0,a,\theta) + 1_{\{ x\in\Xi_{21} \}} \sum_{\vartheta\in\varTheta} V(\vartheta w, n+1, 0,0,\theta) \hat{f}_B(\vartheta,n)\\
&+ 1_{\{ x\in\Xi_{22}\}} V(0,0,0,0,\theta) +  1_{\{ x\in\Xi_{23}\}}V(0,0,0,0,\theta).
\end{align*}
The expression of $\mathfrak{T}V(x)$ is:
\begin{align*}
\mathfrak{T}V(x) =& \inf_{a\in A(x)} \left\{ C^i(x,a) + QV(x,a) \right\}\\
\approx & 1_{\{ x\in\Xi_{\mathrm{end}} \}} V(w,n,\sigma,d,\theta) + \inf_{a\in A(x)} \{ 1_{\{x\in\Xi_1 \}} (V(w,n,0,a,\theta) + C_{\mathrm{isp}}) \}\\
&+ 1_{\{ x\in\Xi_{21} \}} \left(\sum_{\vartheta\in\varTheta} V(\vartheta w, n+1, 0,0,\theta) \hat{f}_B(\vartheta,n) + C_1(x)\right)\\
&+ 1_{\{ x\in\Xi_{22} \}} \left(  V(0,0,0,0,\theta) + C_2 \right)\\
&+ 1_{\{ x\in\Xi_{23} \}} \left( V(0,0,0,0,\theta) + C_3 \right).
\end{align*}

\subsubsection{Computation of $\mathfrak{B}$}
Since we can not directly compute $\mathfrak{B}V$ on $\bar{X}$, we can approximate it on a grid of the state space. In order to approximate $\mathfrak{B}V(x)$ for any $x\in \bar{X}$, we divide $\mathfrak{B}V(x)$ into two parts and compute the value functions of continuous operating cost and impulse control cost separately, that is $\mathfrak{B}V(x) = G(V,x) + H(V,x)$, where 
\begin{align*}
G(V,x) &= \int_{[0,t^*(x))} e^{-(K+\rho)t}\mathfrak{R}V(\phi(x,t)) dt,\\
H(V,x) &= e^{-(K+\rho)t^*(x)} \mathfrak{T}V(\phi(x,t^*(x))).
\end{align*}

Denote the time interval by $\Delta t$. For any $x=(w,n,\sigma,d,\theta)\in X$, define $s^*(x) = \lfloor \frac{t^*(x)}{\Delta t}\rfloor$. For any $j\in \{ 0,\cdots, n^*(x)-1\}$, set $\phi_j(x,t) = \phi(x,j\Delta t)$ and $\phi(x,t^*(x)) = (\varphi(x,t^*(x)),n,\sigma+t^*(x),d,\theta+t^*(x))$, where $t^*(x)$ is the time when the flow hits the boundary starting from $x$. The above $G$ can be approximated by the classical trapezoidal formula for calculating the integral:
\[ G(V,x) \approx \frac{\Delta t}{2} \mathfrak{R}V(x) + \frac{\Delta t}{2} e^{-(K+\rho)t^*(x)} \mathfrak{R}V(\phi(x,t^*(x))) + \sum_{j=1}^{s^*(x)-2} \Delta t e^{-(K+\rho) j\Delta t} \mathfrak{R} V(\phi_j(x,t)). \]
Thus, for $x=(w,n,\sigma,d,\theta)\in X$ we have
\begin{equation*}
\begin{aligned}
G(V,x) \approx & \frac{\Delta t}{2} \left[ C_g + 1_{\{ w \ge \zeta \}} C_{\mathrm{grd}} (w-\zeta) + \eta(x) \sum_{\varpi\in \varPi} V(w+ \varpi,n,\sigma,d,\theta) \hat{f}_{\mathrm{IG}} \left(\varpi,\frac{1}{\eta(x)}\right) \right]\\
&+ \frac{\Delta t}{2} e^{-(K+\rho)t^*(x)} \left[ C_g + 1_{\{ \varphi(x,t^*(x)) \ge \zeta \}} C_{\mathrm{grd}}(\varphi(x,t^*(x)) - \zeta) \vphantom{\sum_{i=1}^\infty} \right. \\
& \left. \vphantom{\sum_{i=1}^\infty} + \eta(\phi(x,t^*(x))) \sum_{\varpi\in\varPi} V(\varphi(x,t^*(x))+ \varpi,n,\sigma+t^*(x),d,\theta+t^*(x)) \hat{f}_{\mathrm{IG}}\left(\varpi,\frac{1}{\eta(\phi(x,t^*(x)))}\right) \right]\\
&+ \sum_{j=1}^{s^*(x)-2} \Delta t e^{-(K+\rho) j\Delta t} \left[ C_g + 1_{\{ \varphi(x,j\Delta t) \ge \zeta \}} C_{\mathrm{grd}} (\varphi(x,j\Delta t) - \zeta) +\vphantom{\sum_{i=1}^\infty} \right.\\
&\left. \vphantom{\sum_{i=1}^\infty} + \eta(\phi(x,j\Delta t)) \sum_{\varpi\in\varPi} V(\varphi(x,j\Delta t) + \varpi,n,\sigma + j\Delta t, d, \theta + j\Delta t ) \hat{f}_{\mathrm{IG}}\left(\varpi,\frac{1}{\eta(\phi(x,j\Delta t))}\right) \right].
\end{aligned}
\end{equation*}

For $H(V,x)$, we can determine it according to the different boundaries hit by $\phi(x,t^*(x))$. More precisely,
\begin{itemize}
	\item If $\phi(x,t^*(x)) \in \Xi_1$, i.e. $\sigma+t^*(x) = T_{\mathrm{isp}}$, then
	\[ H(V,x) = e^{-(K+\rho)t^*(x)}\inf_{a\in A(x)} \left\{ V(\varphi(x,t^*(x)),n,0,a,\theta + t^*(x)) + C_{\mathrm{isp}} \right\}. \]
	\item If $\phi(x,t^*(x))\in \Xi_{21}$, i.e. $\sigma+t^*(x) = T_{\mathrm{soj}}$, $d=1$  and $\varphi(x,t^*(x)) < M$, then
	\[ H(V,x) \approx e^{-(K+\rho)t^*(x)} \left[ \sum_{\vartheta\in\varTheta} V(\vartheta \varphi(x,t^*(x)), n+1,0,0,\theta+t^*(x)) \hat{f}_B(\vartheta,n) + C_1(x)\right].  \]
	\item If $\phi(x,t^*(x)) \in \Xi_{22}$, i.e. $\sigma+t^*(x) = T_{\mathrm{soj}}$ and $d=2$, then
	\[ H(V,x) = e^{-(K+\rho)t^*(x)} \left[ V(0,0,0,0,\theta +t^*(x)) + C_2\right]. \]
	\item If $\phi(x,t^*(x)) \in \Xi_{23}$, i.e. $\sigma+t^*(x) = T_{\mathrm{soj}}$,   $d=1$ and $\varphi(x,t^*(x)) \ge M$, then
	\[ H(V,x) = e^{-(K+\rho)t^*(x)} \left[ V(0,0,0,0,\theta+t^*(x)) + C_3\right]. \]
	\item If $\phi(x,t^*(x)) \in \Xi_{\mathrm{end}}$, i.e. $\theta + t^*(x) = T_{\mathrm{end}}$, then
	\[ H(V,x) = e^{-(K+\rho)t^*(x)} V(\varphi(x,t^*(x)),n,\sigma+t^*(x),d,T_{\mathrm{end}}). \]
\end{itemize}

To sum up, for any $x=(w,n,\sigma,d,\theta)\in \bar{X}$. The expression of $\mathfrak{B}V(x)$ can be written as: 
\begin{align*}
\mathfrak{B}& V(x) \approx \frac{\Delta t}{2} \left[ C_g + 1_{\{ w > \zeta \}} C_{\mathrm{grd}} (w-\zeta) + \eta(x) \sum_{\varpi\in \varPi} V(w+ \varpi,n,\sigma,d,\theta) \hat{f}_{\mathrm{IG}} \left(\varpi,\frac{1}{\eta(x)}\right) \right]\\
&+ \frac{\Delta t}{2} e^{-(K+\rho)t^*(x)} \left[C_g + 1_{\{ \varphi(x,t^*(x)) > \zeta \}} C_{\mathrm{grd}}(\varphi(x,t^*(x)) - \zeta) \vphantom{\sum_{i=1}^\infty} \right. \\
& \left. \vphantom{\sum_{i=1}^\infty} + \eta(\phi(x,t^*(x))) \sum_{\varpi\in\varPi} V(\varphi(x,t^*(x))+ \varpi,n,\sigma+t^*(x),d,\theta+t^*(x)) \hat{f}_{\mathrm{IG}}\left(\varpi,\frac{1}{\eta(\phi(x,t^*(x)))}\right) \right]\\
&+ \sum_{j=1}^{s^*(x)-2} \Delta t e^{-(K+\rho) j\Delta t} \left[ C_g + 1_{\{ \varphi(x,j\Delta t) > \zeta \}} C_{\mathrm{grd}} (\varphi(x,j\Delta t) - \zeta) +\vphantom{\sum_{i=1}^\infty} \right.\\
&\left. \vphantom{\sum_{i=1}^\infty} + \eta(\phi(x,j\Delta t)) \sum_{\varpi\in\varPi} V(\varphi(x,j\Delta t) + \varpi,n,\sigma + j\Delta t, d, \theta + j\Delta t ) \hat{f}_{\mathrm{IG}}\left(\varpi,\frac{1}{\eta(\phi(x,j\Delta t))}\right) \right]\\
&+ 1_{\{ \sigma+t^*(x) = T_{\mathrm{isp}}\}} e^{-(K+\rho)t^*(x)}\inf_{a\in A(x)} \left\{ V(\varphi(x,t^*(x)),n,0,a,\theta + t^*(x)) + C_{\mathrm{isp}} \right\}\\
&+ 1_{\{\sigma+t^*(x) = T_{\mathrm{soj}},d=1,\varphi(x,t^*(x))<M \}} e^{-(K+\rho)t^*(x)} \left[ \sum_{\vartheta\in\varTheta} V(\vartheta \varphi(x,t^*(x)), n+1,0,0,\theta+t^*(x)) \hat{f}_B(\vartheta,n) + C_1(x)\right]\\
&+ 1_{\{\sigma+t^*(x) = T_{\mathrm{soj}},d=2\}}e^{-(K+\rho)t^*(x)} \left[ V(0,0,0,0,\theta +t^*(x)) + C_2\right]\\
&+ 1_{\{\sigma+t^*(x) = T_{\mathrm{soj}},d=1,\varphi(x,t^*(x))\ge M\}}e^{-(K+\rho)t^*(x)}\left[ V(0,0,0,0,\theta+t^*(x)) + C_3\right]\\
&+ 1_{\{ \theta+t^*(x) = T_{\mathrm{end}} \}}  e^{-(K+\lambda)t^*(x)} V(\varphi(x,t^*(x)),n,\sigma+t^*(x),d,T_{\mathrm{end}}).
\end{align*}
and $\mathfrak{B}V(\Delta) = 0$.

\section{Application and simulation}\label{section4}

We apply the preceding results to a component coating maintenance model as an illustration. Multi-layer protective coatings are applied to the surface of materials to resist corrosion of those affected by the environment. Due to natural evaporation or wear, the protective coating's thickness will gradually diminish over time. Accidents such as collisions and chemical liquid corrosions, may potentially result in a decrease in coating thickness or even direct peeling. The substance is immediately exposed to the air when the covering has entirely worn off. Sometimes this exposure might lead to structural failures of components, which have serious safety consequences. Preventive maintenance is necessary to avoid material failures, such as patching anti-corrosion paint, replacing parts, etc.
In this model, repairing protective paint can be regarded as imperfect maintenance, replacing new parts can be considered as corrective maintenance. For researches on corrosion modeling and maintenance, interested readers could refer to \cite{Fatima1,LiYX1}.

\begin{figure}[htbp]
	\centering
	\includegraphics[scale=0.5]{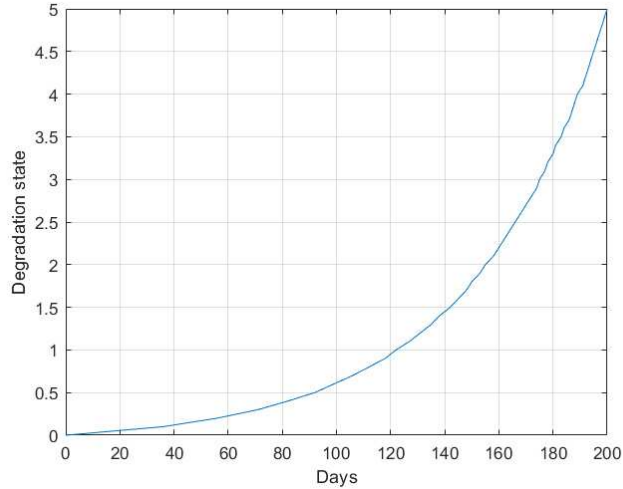}
	\caption{\it The relationship between natural loss of coating thickness $w$ and days.}
	\label{AS-1}
\end{figure}

We assume that the time unit of simulation is day, and the total simulation time is one year, which means $T_{\mathrm{end}} = 365$. Assume that the thickness of the coating is 5 mm.
The relationship between natural loss of thickness and days can be determined by ODEs based on different materials and environments, see \cite{Fatima1,Arya1}. It can also be modeled by statistic methods, such as regression analysis based on past data, see \cite{Panchenko1,Panchenko2}. For the sake of simplicity in the simulation, we assume that the thickness loss of the coating due to natural corrosion increases exponentially. The exponential model is quite prevalent in degradation modeling, such as \cite{Karanci1,Panchenko2}. The illustrative equation of loss of thickness and days is given by $w = \frac{1}{10}\exp(\frac{\ln 51}{200}t)-0.1$. 
This indicates that the time required for the coating to completely vanish due to natural corrosion is 200 days. For computational convenience, we discretize the value of $w$, with interval of 0.1 mm, which means that $w\in\{ 0,0.1,\cdots,5\}$ and there are 51 degradation states in total. Based on the above formula, the sojourn time of the system at each thickness under natural corrosion can be calculated. The relationship between $w$ and days is displayed in Figure \ref{AS-1}.

Assume that the jump intensity of random shocks is $\eta(x) = \frac{w+1}{60}$. This means that the maximum average interval between random shocks is 60 days, and the minimum average interval is 10 days. Since $\eta$ is related with $w$, this reveals that the deterioration status is inversely correlated with the length of time between random shocks. 
Suppose that the inspection interval of the maintenance personnel is 20 days, which means the subsequent observation would occur 20 days after the previous inspection or maintenance.  If the maintenance staff decides to plan a maintenance action during inspection time, they will need to wait for a period of time to carry out maintenance due to the need to prepare necessary equipments and materials. 
We assume that the waiting interval is 5 days, which means $T_{\mathrm{soj}} = 5$. It is easy to calculate that the maximum number of maintenance in this setting is 14 times. The inspection cost is $C_{\mathrm{isp}} = 1$. The fixed operating cost per day is $C_g = 0$. The threshold $\zeta = 4$. When the degradation state of the unit exceeds 4, additional cost $C_{\mathrm{grd}} = w-3$ will incur. So the continuous operation cost is $C^g = 1_{\{w\ge\zeta\}}(w-3)$. The cost of imperfect maintenance is $C_1 = \lfloor w \rfloor + n$, where the fixed maintenance cost $c_1=0$. The cost of corrective maintenance is $C_2 = 10$. When the planned action is imperfect maintenance and the system is in the failure state at maintenance time, the cost of performing a corrective maintenance is $C_3 = 20$. No maintenance cost will be incurred if no maintenance action is performed.

We consider the discount model. It is very significant for decision-makers who want to save costs as much as possible to be able to calculate the cost that will be paid in the future and the principal that will be prepared at the present time according to the discount factor $\rho$. With the help of the discount model, decision-makers can reduce the principal reserve and free up more money for other investment activities by depositing the principal with a fixed interest rate in the bank in advance, then withdrawing it and paying it when the cost arises. The classic total cost model is obtained when the discount factor $\rho$ approaches zero.

According to the result in the above subsections, we compute (\ref{thm1-1}) iteratively. Take $\varepsilon = 0.01$, when $\norm{W_{k+1}-W_{k}}_\infty < \varepsilon$, stop the iteration, then we obtain the approximate optimal matrix $W$. Since we need to make a bijection between the five tuple $x=(w,n,\sigma,d,\theta)$ and the matrix $W$, interested readers can refer to the method of building matrix in \cite{Dufour1} Section 4.2. If the discount factor $\rho = 0.001$, based on the above data, the stopping limit can be met in about 30 iterations. Due to the complexity of the problem, the computation procedure is quite time-consuming. 
Many factors contribute to the complexity of computation, such as the discount factor $\rho$, number of imperfect maintenance $n$, improvement factor $\vartheta$ and the solvability of the deterministic ODEs of the flow $\phi$, thus the computational complexity of the algorithm is still an open question. 
Various methods are possible to enhance the calculation efficiency, such as quantization method used in \cite{Dufour3} and $\varepsilon$-optimal algorithm in \cite{Dufour4}.
The simulation is performed on an Intel i7-6700HQ CPU, and the computing time for discount factor $\rho = 0.001$ to achieve the approximate optimal policy is about 6 hours.

\subsection{The simulation of PDMP paths}
\begin{figure}[htbp]
	\centering
	\subfigure{
		\begin{minipage}{8cm}
			\centering
			\includegraphics[scale=0.4]{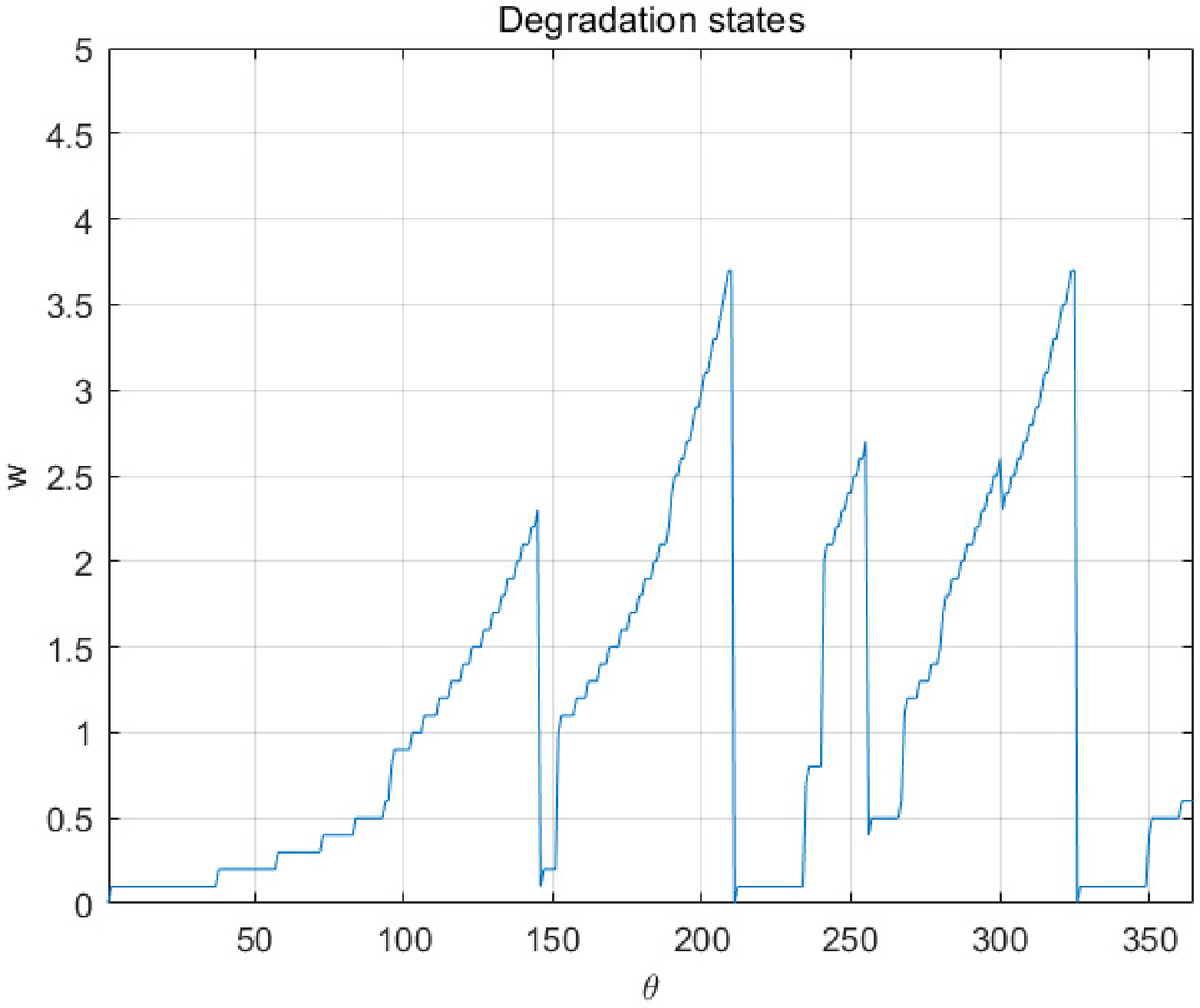}
		\end{minipage}
	}
	\subfigure{
		\begin{minipage}{8cm}
			\includegraphics[scale=0.4]{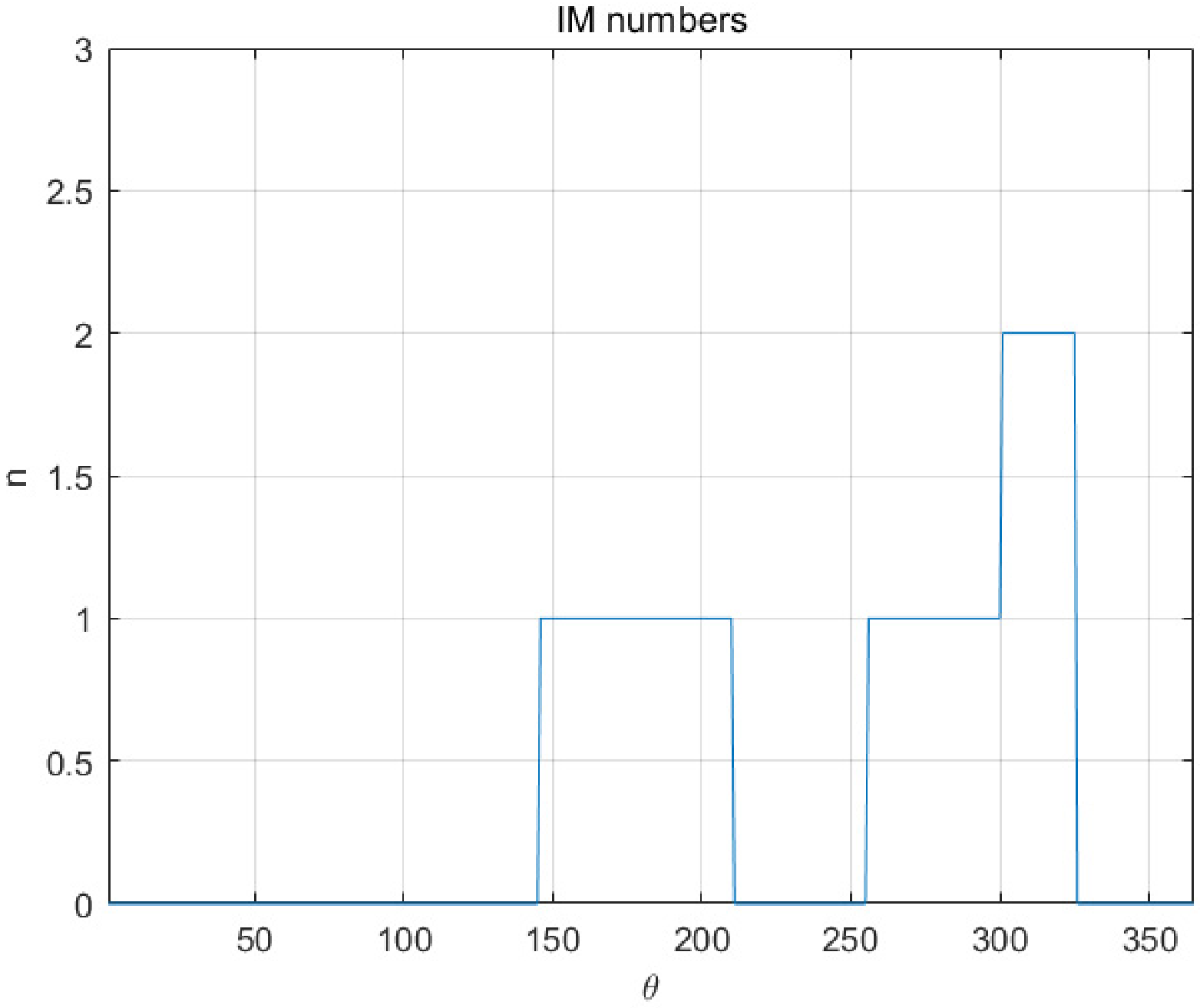}
		\end{minipage}
	}
	
	\subfigure{
		\begin{minipage}{8cm}
			\centering
			\includegraphics[scale=0.4]{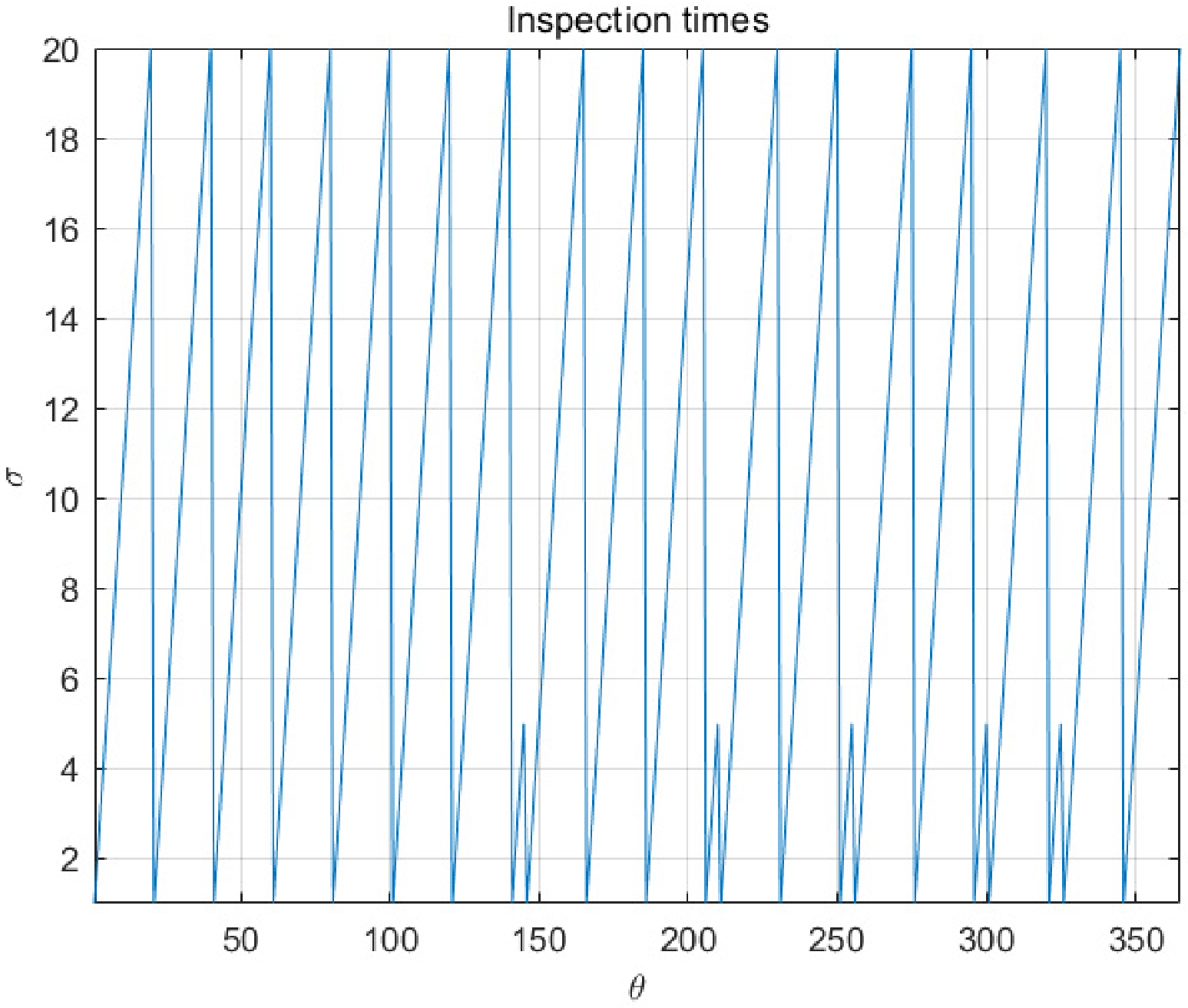}
		\end{minipage}
	}
	\subfigure{
		\begin{minipage}{8cm}
			\includegraphics[scale=0.4]{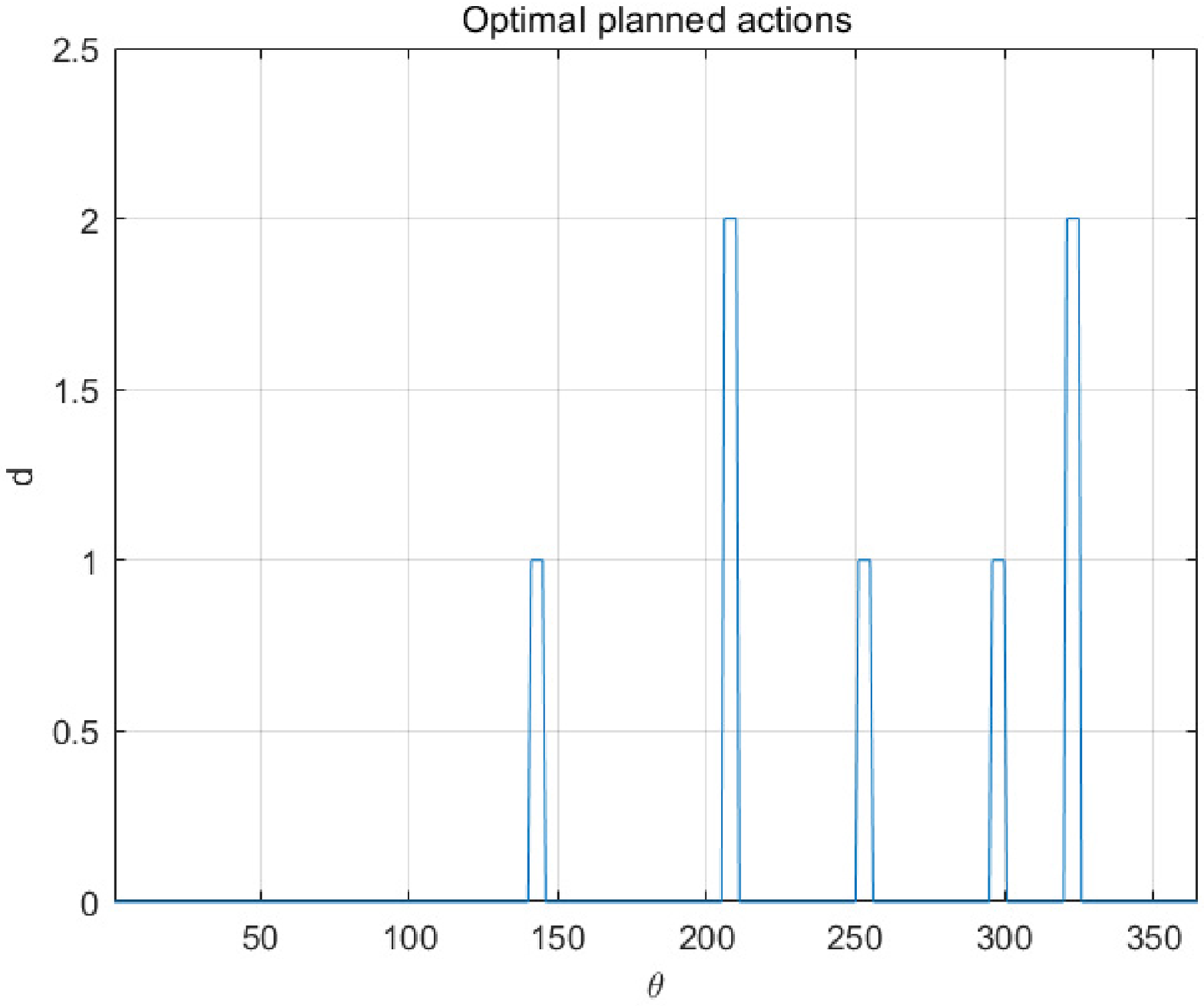}
		\end{minipage}
	}
	\caption{\it A simulated PDMP path of IMM. The above four figures show the degradation state $w$ of the system, the execution times of imperfect maintenance $n$, the change of observation time $\sigma$, and the planned optimal action $d$ at each time under this path respectively.}
	\label{AS-2}
\end{figure}

When the discount factor $\rho = 0.001$, we let the system begin from state $x=(0,0,1,0,1)$ and simulate a path of our imperfect maintenance model (IMM). The result is shown in Figure \ref{AS-2}.

\begin{figure}[htbp]
	\centering
	\begin{minipage}{8cm}
		\centering
		\includegraphics[scale=0.4]{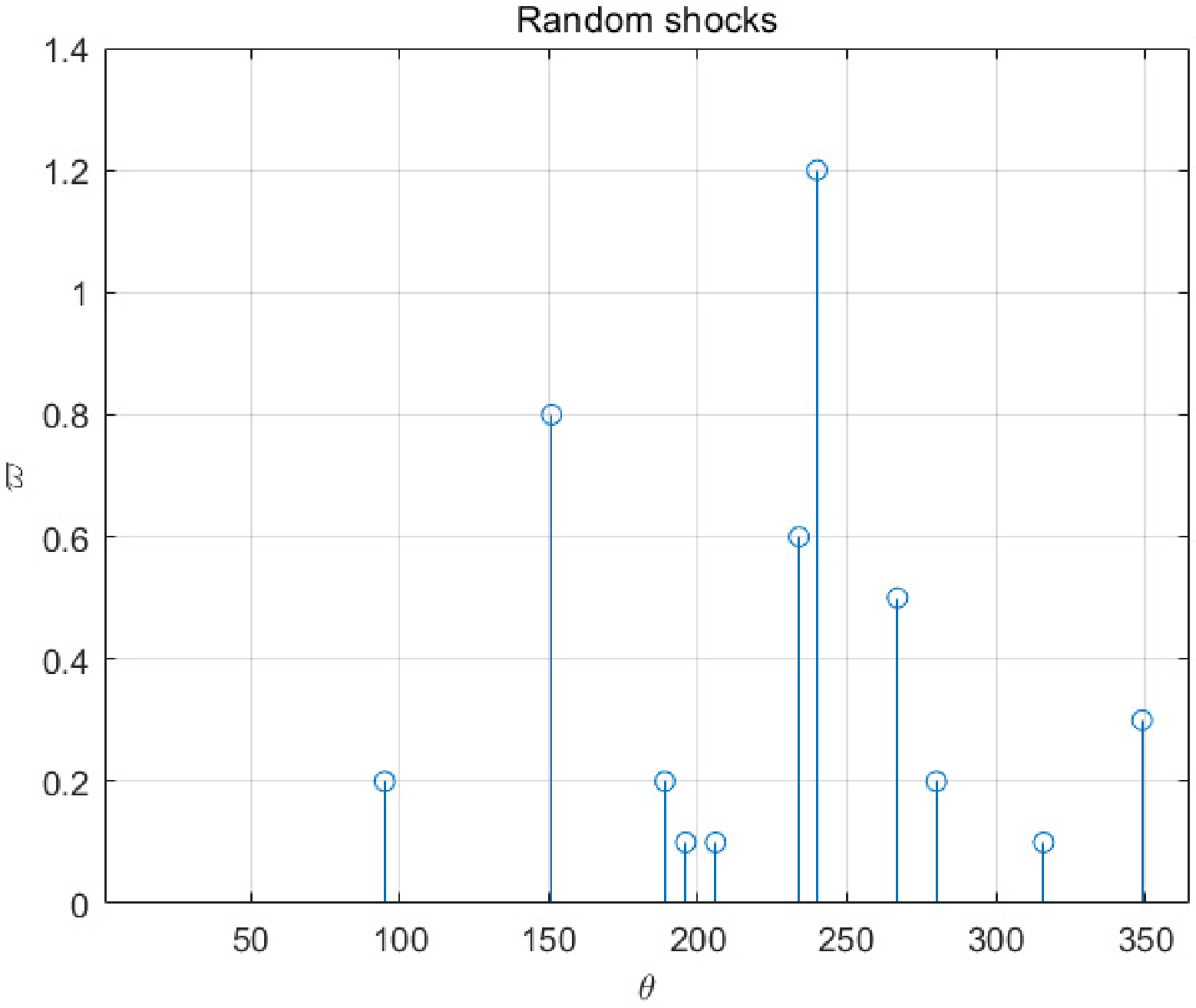}
	\end{minipage}
	\begin{minipage}{8cm}
		\includegraphics[scale=0.4]{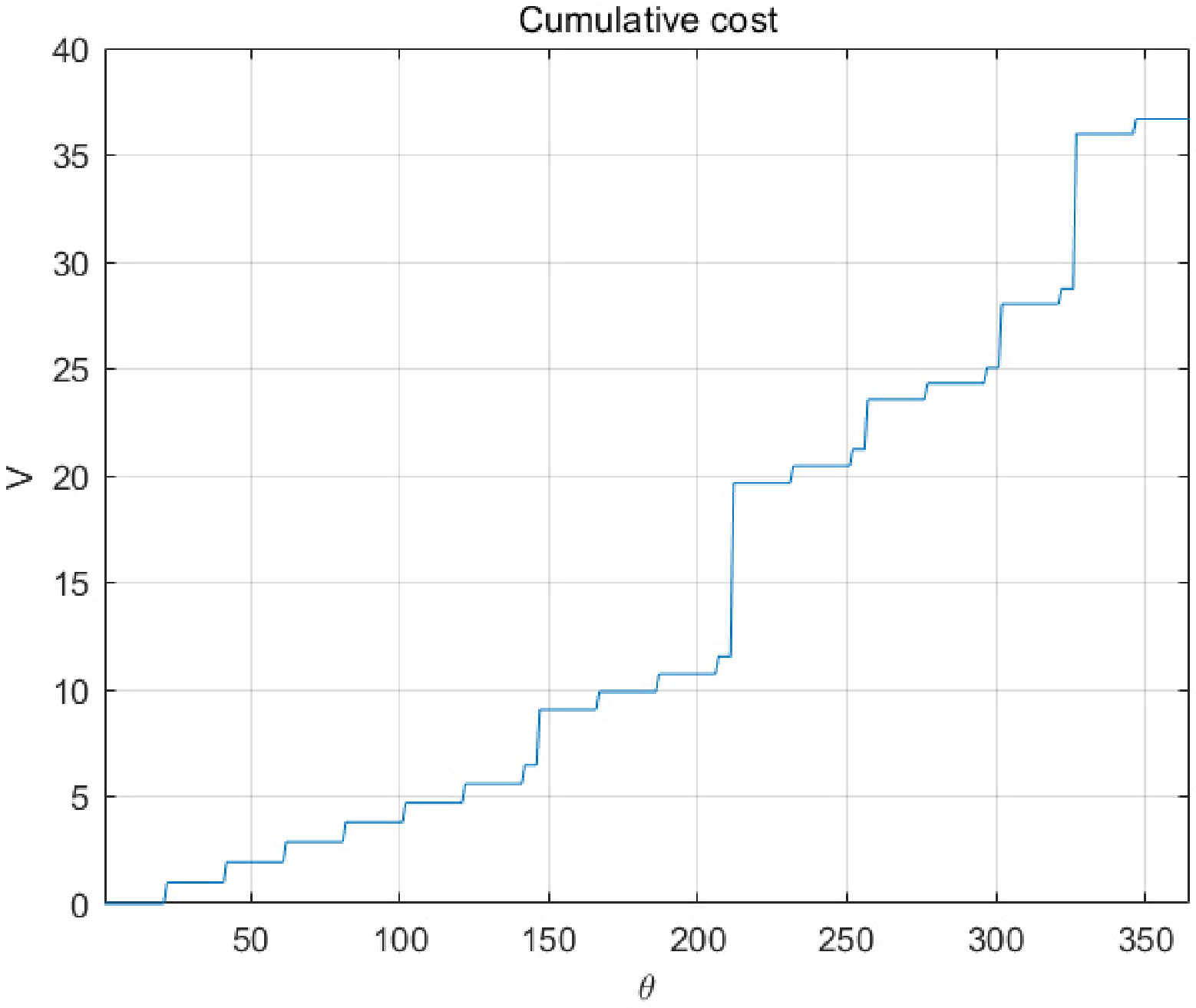}
	\end{minipage}
	\caption{\it The left figure is the time of random shocks and damage increment $\varpi$ of a simulation path of PDMP model, the right figure is the cumulative discounted cost $V$ of that path.}
	\label{AS-3}
\end{figure}

The four figures in Figure \ref{AS-2} show the changes of simulated degradation states, executed imperfect maintenance times, change of inspection time and optimal planned action respectively. The first random shock happened at $\theta=95$, it increased the degradation state from 0.7 to 0.9. In the first six inspection times, $\theta = 20,40,60,80,100,120$, the decision-maker made the decision of no maintenance. In the 7th inspection time, $\theta = 140$, the degradation state is 2.1, the decision-maker decided to schedule an imperfect maintenance for the first time. When $\theta = 145$, the imperfect maintenance was performed and the degradation state decreased from 2.3 to 0.1. The second random shock happened at $\theta = 151$ and increased the degradation state from 0.2 to 1.0. In the 10th inspection time, $\theta = 205$, the decision-maker scheduled a corrective maintenance. And in the 12th and 14th inspection time, the decision-maker scheduled an imperfect maintenance plan twice in a row. The time of random shocks and damage increments $\varpi$ is shown in the left figure of Figure \ref{AS-3}. It can be observed that the frequency of random shocks will increase as the degree of degradation increases. At $\theta = 151$ and 240, the random shocks caused two major damage to the system, with increasing damage $\varpi = 0.8$ and 1.2 respectively. Fortunately, the system was in a state of low damage when the two random shocks occurred, and the two random shocks with large damage increment did not make the system enter the failure state. The system was in the highest degradation state when $\theta = 209$, but corrective maintenance was performed in $\theta = 210$ and reset the degradation state of the system to zero. The right figure of Figure \ref{AS-3} shows the cumulative discounted cost of the system. The total discounted cost of this simulation is 36.68.

As a comparison, we simulate two models with fixed maintenance strategy. The first model is the most basic maintenance model, that is, the decision-maker does not plan to repair until he observes that the system is in the failure state and the only maintenance action adopted is corrective maintenance. We call this model as corrective maintenance model (CMM). The second model is a common condition-based preventive maintenance model. In this model, we set two thresholds $\xi_1,\xi_2$. When the degradation state is less than $\xi_1$, no maintenance action is performed. When the degradation state is between $\xi_1$ and $\xi_2$, imperfect maintenance action is scheduled. When the degradation state is greater than $\xi_2$, corrective maintenance action is planned. Thus the maintenance action can be determined through the following function:
\[ a(w) = \begin{cases}
0, & 0\le w < \xi_1,\\
1, & \xi_1 \le w < \xi_2,\\
2, & \xi_2 \le w \le 5.
\end{cases} \]
We call this kind of model as threshold maintenance model (TMM). Of course, if an imperfect maintenance is planned and the system is in failure state during the maintenance, the corrective maintenance will be carried out immediately and cost $C_3$ will incur. We also simulate a PDMP path for these two models. Other parameters of the two models are the same as before, the thresholds of TMM is set as $\xi_1 = 2.0$, $\xi_2 = 4.0$. The results are shown in the following figures.

\begin{figure}[htbp]
	\centering
	\subfigure{
		\begin{minipage}{8cm}
			\centering
			\includegraphics[scale=0.4]{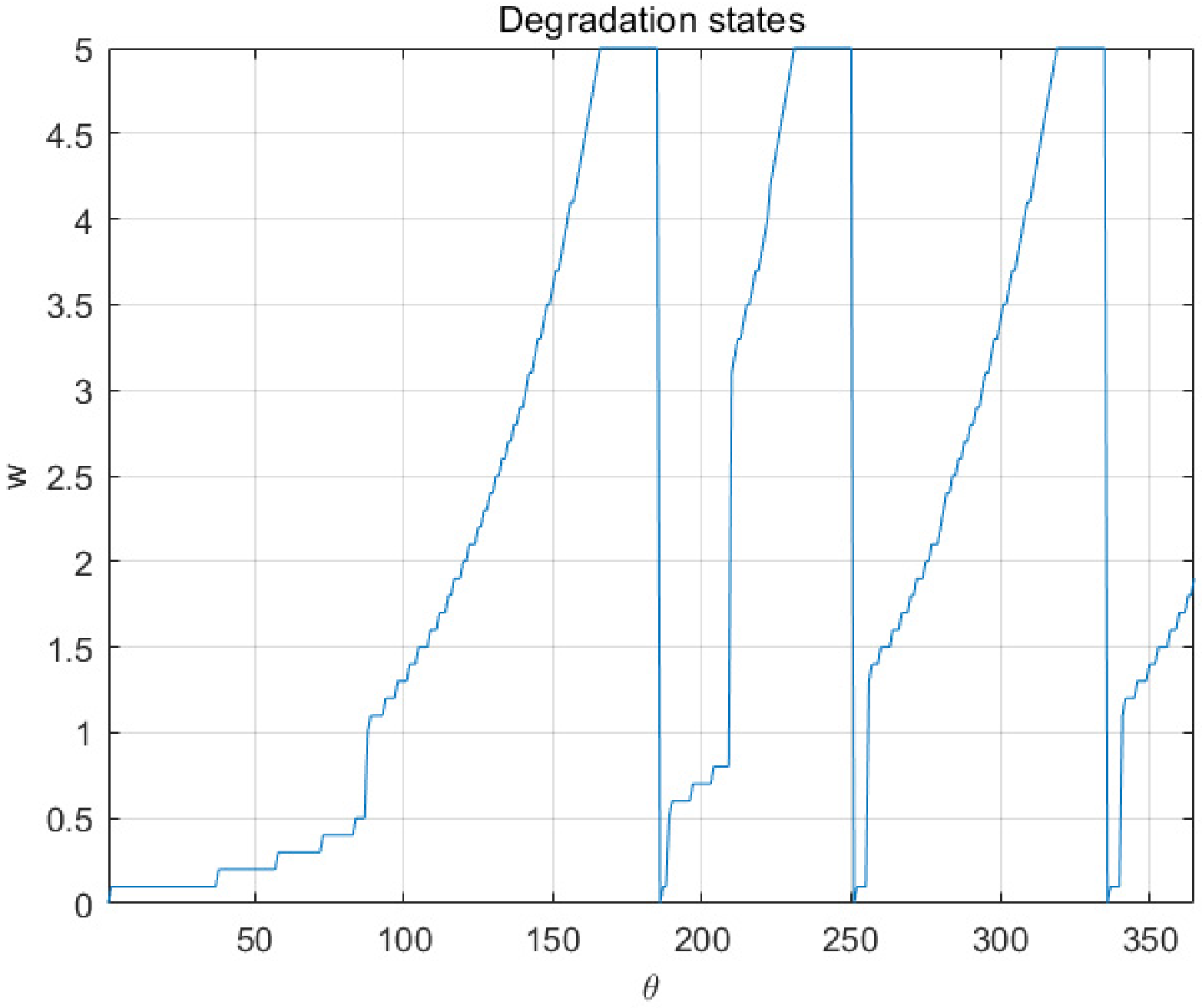}
		\end{minipage}
	}
	\subfigure{
		\begin{minipage}{8cm}
			\includegraphics[scale=0.4]{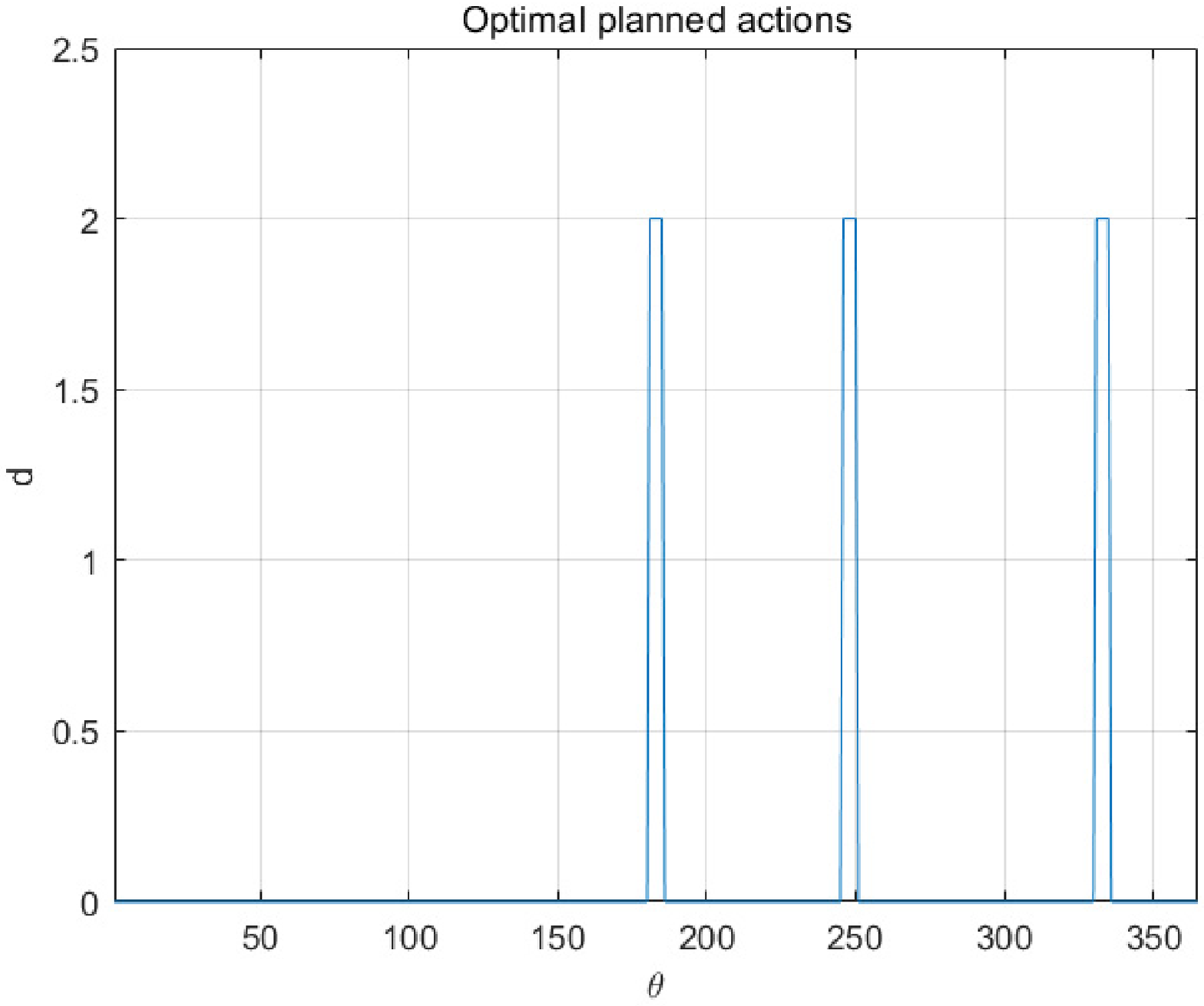}
		\end{minipage}
	}
	
	\subfigure{
		\begin{minipage}{8cm}
			\centering
			\includegraphics[scale=0.4]{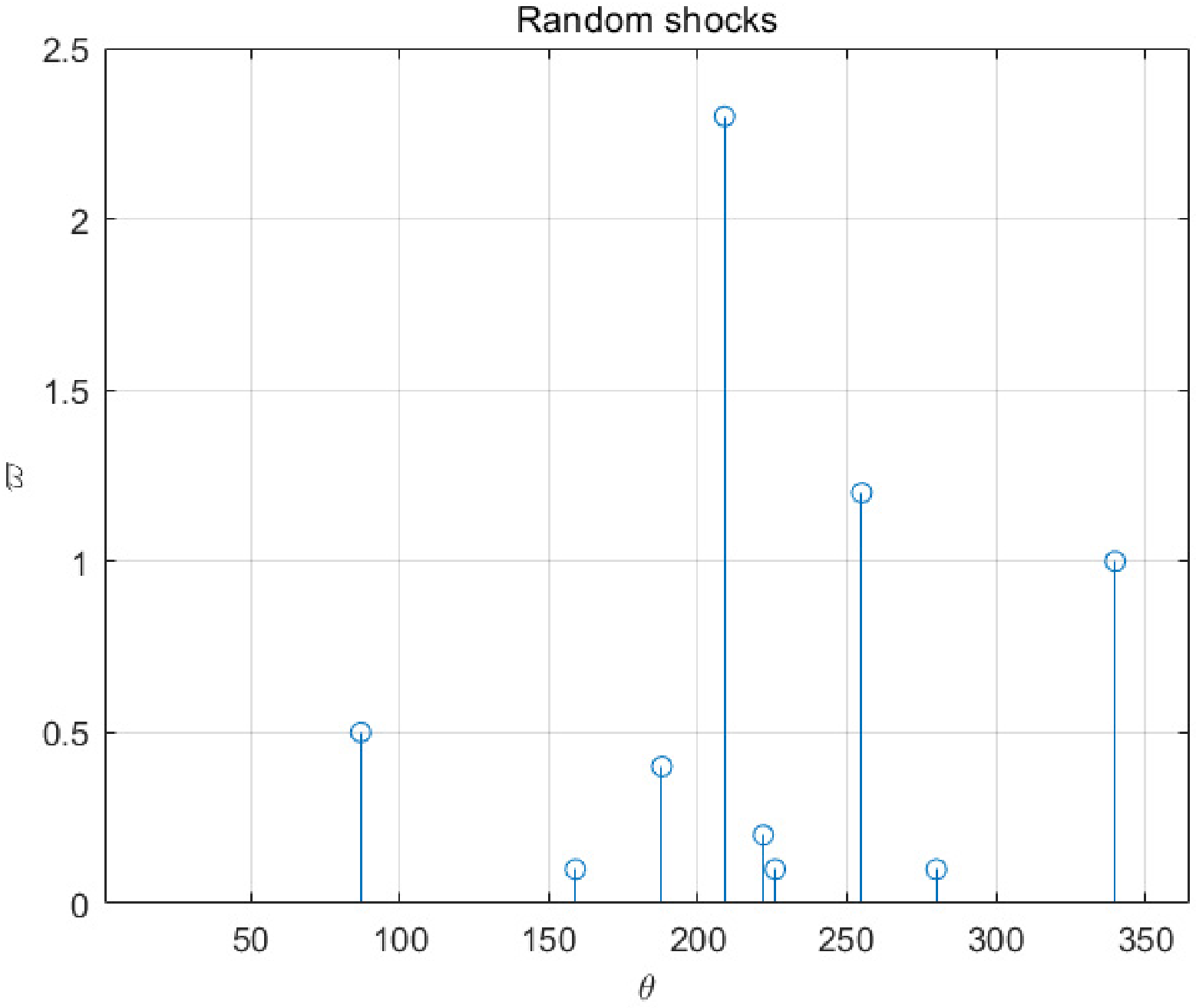}
		\end{minipage}
	}
	\subfigure{
		\begin{minipage}{8cm}
			\includegraphics[scale=0.4]{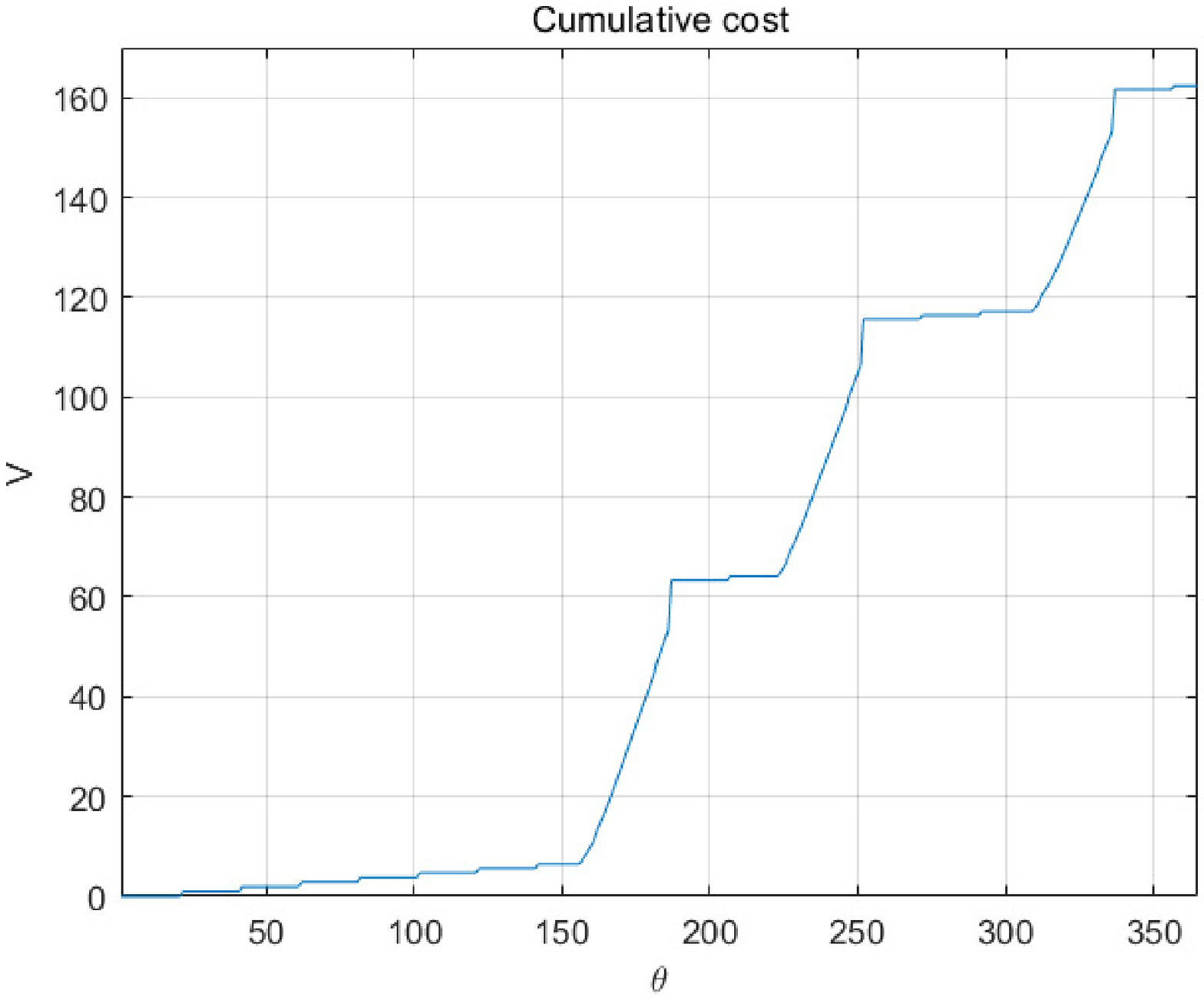}
		\end{minipage}
	}
	\caption{\it A PDMP simulation path of CMM. The above four figures show the degradation state $w$, planned action $d$, the time of random shocks, damage increment $\varpi$, and cumulative discounted cost $v$ respectively.}
	\label{AS-5}
\end{figure}

The simulation result of CMM is shown in Figure \ref{AS-5}. As can been seen that the system failed at $\theta = 166,231$ and 319. The corrective maintenance actions were performed at $\theta = 185,250$ and 335. 
The system remains in the failed state for a considerable amount of time since it waits until it has failed before planning maintenance actions. In this simulation, the overall time of the system in the failed state is 57 days. The total discounted price is 162.36. Because preventive maintenance is not taken into account in CMM, even though the number of maintenance is reduced, operating costs are significantly increased and the time that the system spends in the failed state also increases, which is not favorable for units that must actively avoid the failure state.

\begin{figure}[htbp]
	\centering
	\subfigure{
		\begin{minipage}{8cm}
			\centering
			\includegraphics[scale=0.4]{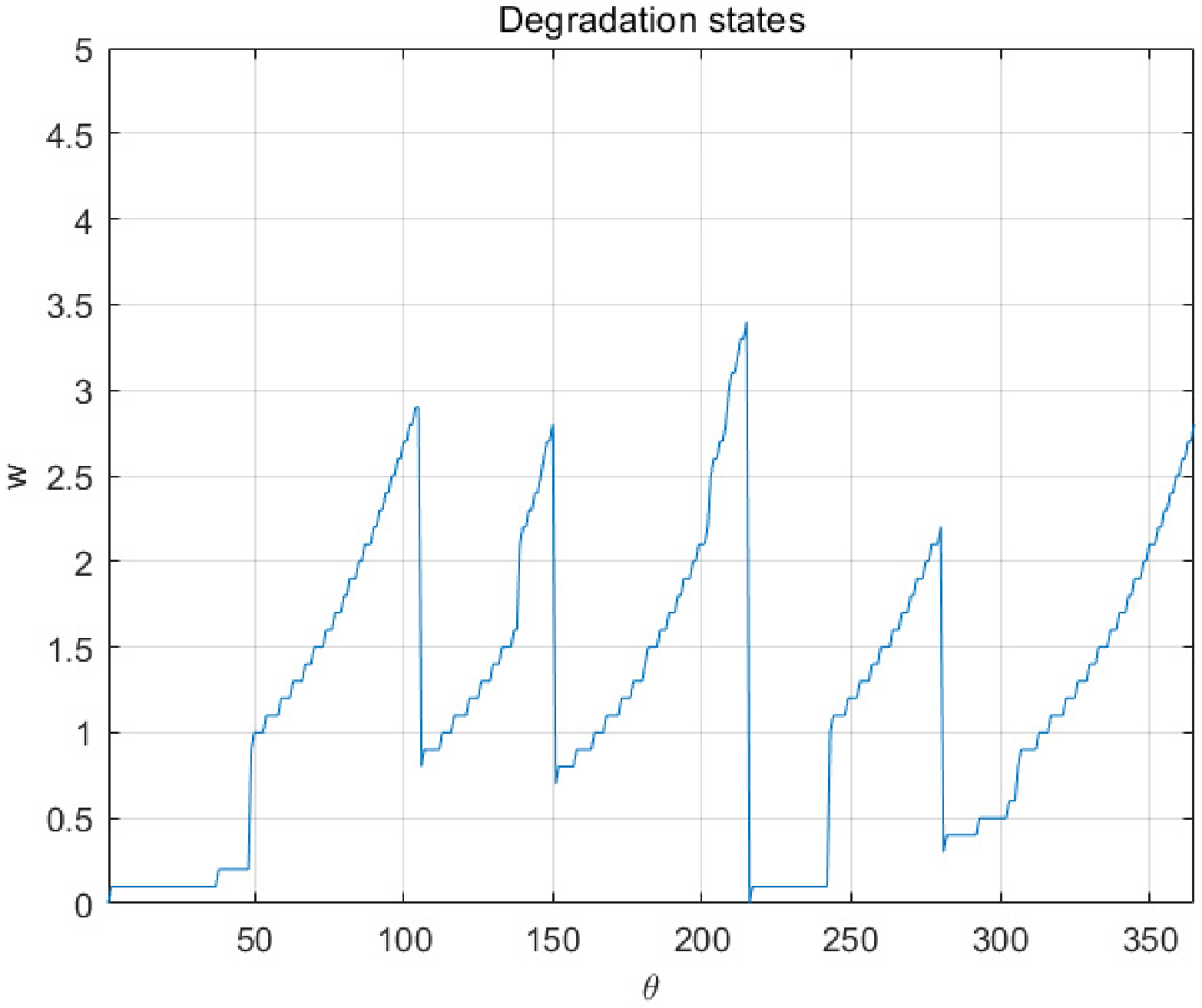}
		\end{minipage}
	}
	\subfigure{
		\begin{minipage}{8cm}
			\includegraphics[scale=0.4]{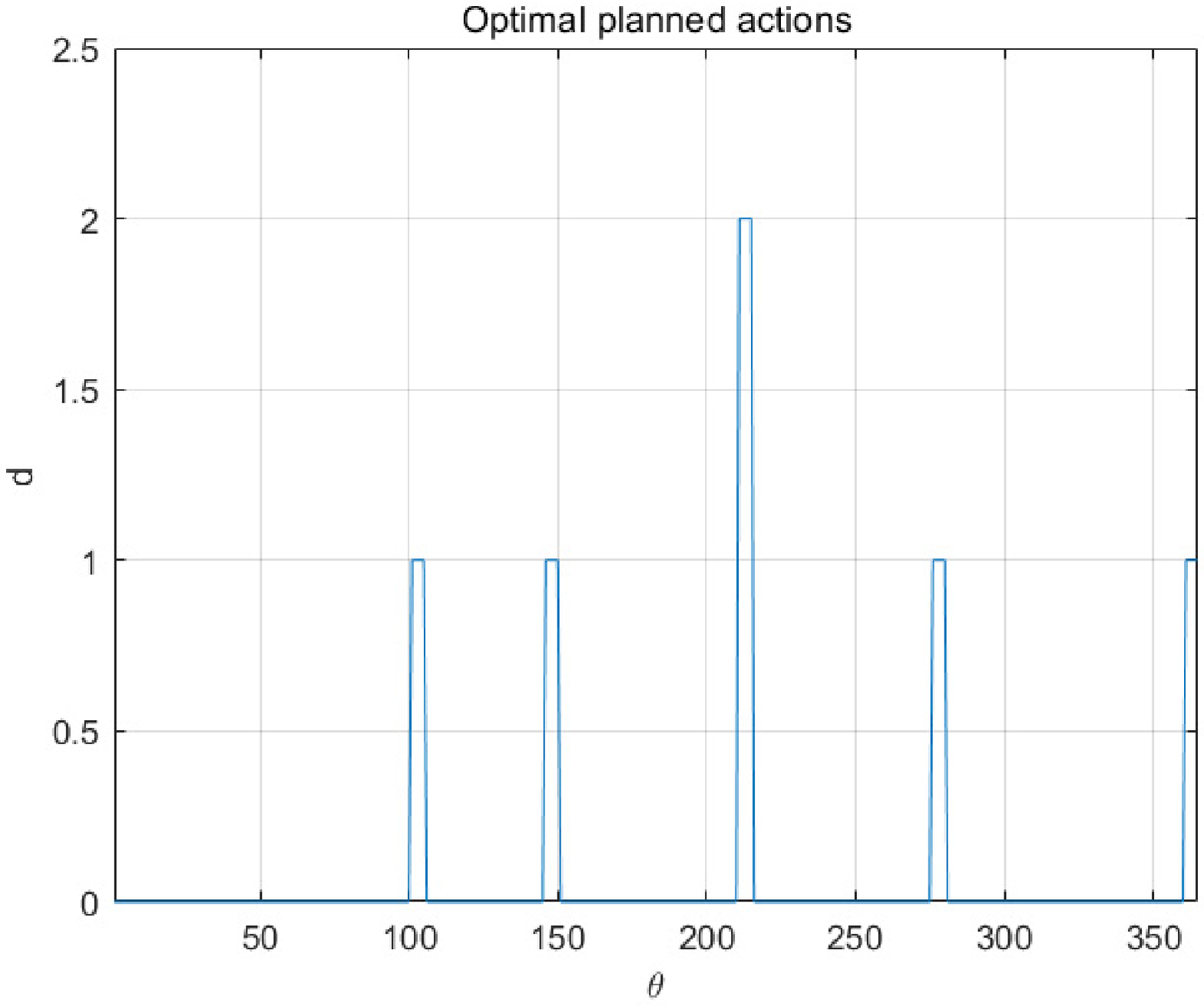}
		\end{minipage}
	}
	
	\subfigure{
		\begin{minipage}{8cm}
			\centering
			\includegraphics[scale=0.4]{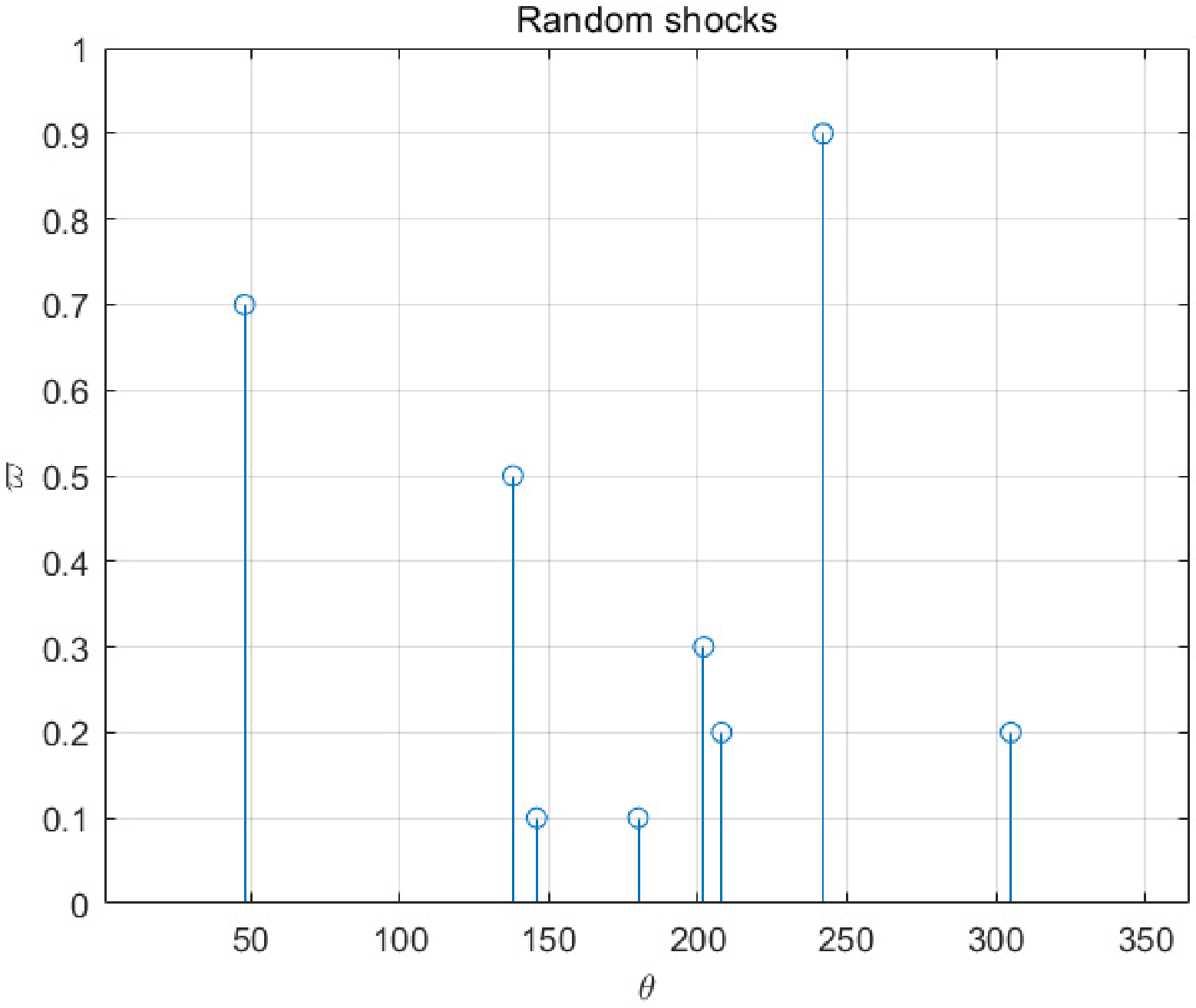}
		\end{minipage}
	}
	\subfigure{
		\begin{minipage}{8cm}
			\includegraphics[scale=0.4]{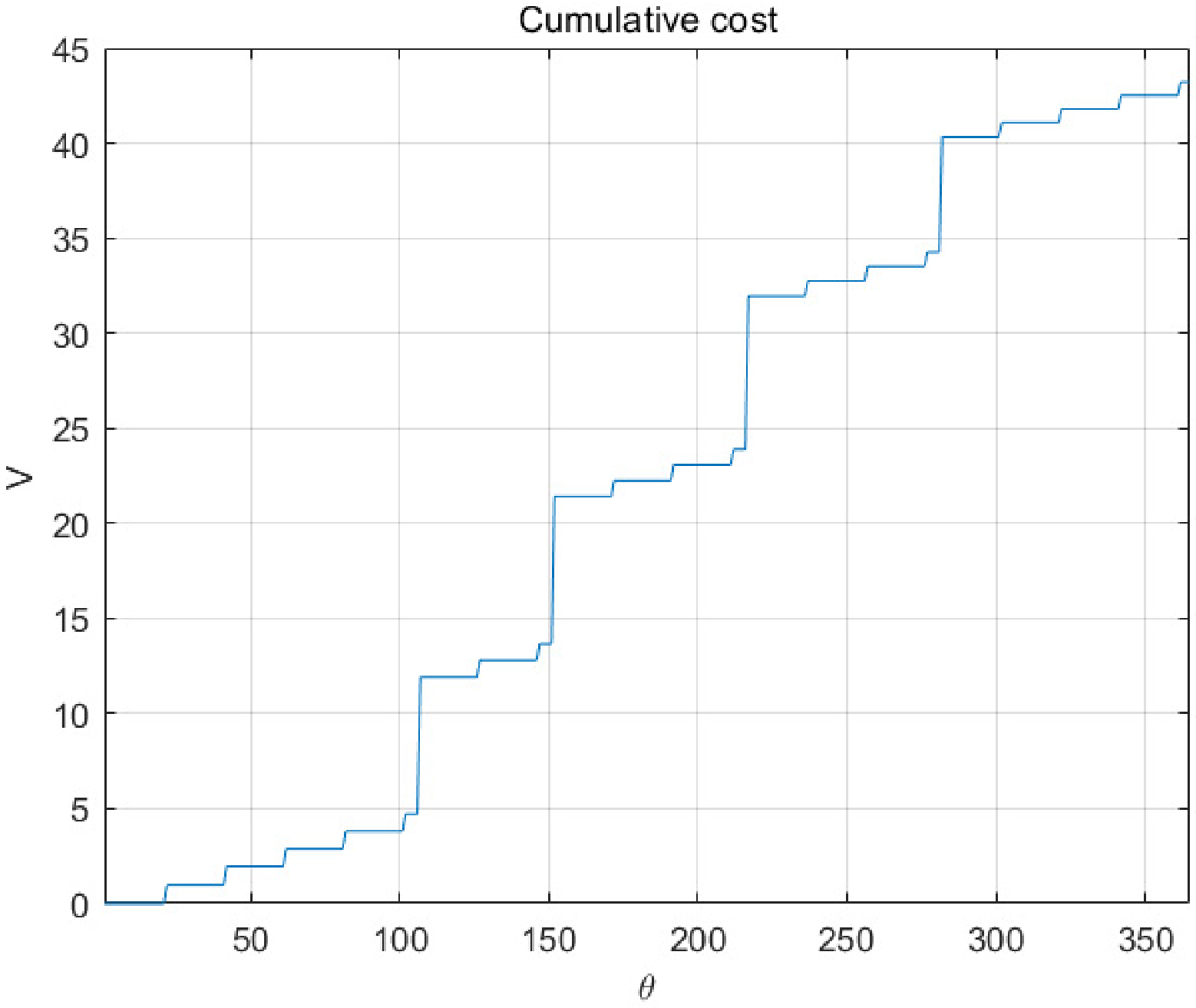}
		\end{minipage}
	}
	\caption{\it A PDMP simulation path of TMM. The above four figures show the degradation state $w$, planned action $d$, the time of random shocks, damage increment $\varpi$, and cumulative discounted cost $V$ respectively.}
	\label{AS-6}
\end{figure}

The simulation result of TMM is shown in Figure \ref{AS-6}. Since we set the maintenance threshold as $\xi_1 = 2.0$, when the degradation state exceeds 2, the decision-maker will schedule a maintenance action. In this simulation, the first random shock happened at $\theta = 48$, the increment of  damage was 0.7, and the degradation state changed from 0.2 to 0.9. The system totally suffered 8 times of shocks. The maximum random damage increment is 0.9. A total of 4 maintenance tasks were carried out, including 3 imperfect maintenance tasks and 1 corrective maintenance task. The total discounted cost is 43.20. It can be seen that the TMM that randomly selected the threshold values has the problem that the planned maintenance task is not timely. The choice of threshold values $\xi_1,\xi_2$ has a crucial impact on the total discounted cost.

\begin{figure}[htbp]
	\centering
	\includegraphics[scale=0.5]{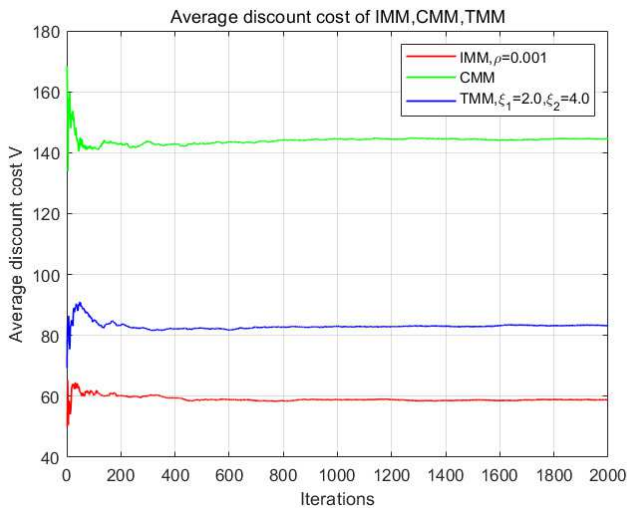}
	\caption{\it Average discounted cost of IMM, CMM, TMM with 2000 simulation paths.}
	\label{AS-7}
\end{figure}

Next, we calculate the average discounted cost of the above three models, that is, each model simulates 2000 paths and calculates the average discounted cost. The results are shown in Figure \ref{AS-7}. The red line, green line and blue line represent IMM, CMM and TMM respectively. It can be seen that the mean values of the three models have converged in 1000 iterations. The average discounted cost of our model is optimal, about 58.57. The second one is the TMM with thresholds $\xi_1=2.0,\xi_2=4.0$, about 83.27. The worst is the CMM, about 144.57, almost 2.5 times of the average discounted cost of IMM. This also shows the necessity of preventive maintenance strategies. In the following subsection, we will discuss the relationship between our model and threshold maintenance model, and find the optimal thresholds combination of TMM.

\subsection{The relationship with threshold maintenance model}

It is a very common kind of maintenance strategy to determine maintenance actions through thresholds. Such maintenance strategies are widely used in preventive maintenance models, such as \cite{Alaswad1,LiYanfu1,Mosayebi1}. In this subsection, we discuss the relationship between our model and the threshold maintenance model.

Generally, the degradation state of the system can be selected as the threshold value. Since the computation of invariant distribution of PDMP is very complex, we could not apply the method used in \cite{Mosayebi1} or \cite{Arismendi1} to calculate the limiting distributions of the system to determine the thresholds. It is worth noting that, the method of calculating the C-K equation in \cite{Arismendi1} is not applicable to our model. This is because in \cite{Arismendi1}, the continuous variable is only time, and the degradation state changes linearly, the random jumps has deterministic transition matrix. In our model, the degradation state is continuous, and the change of degradation could be nonlinear. Our model also has random shocks and the increment  of damage subjects to IG distribution. Imperfect maintenance is also considered. These greatly increases the computational complexity of the C-K equation, so that it is impossible to solve such an equation.

\begin{figure}[htbp]
	\centering
	\includegraphics[scale=0.5]{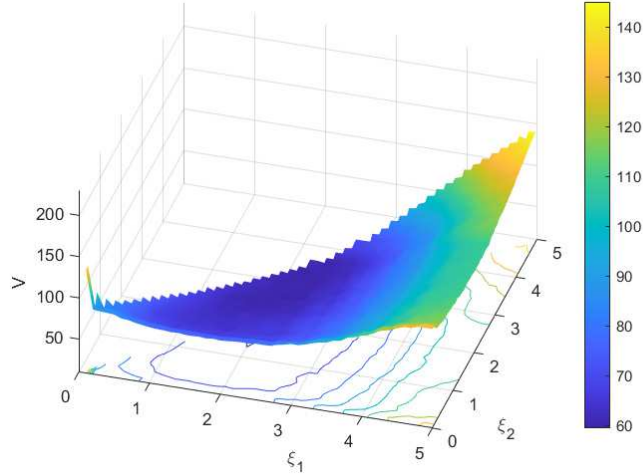}
	\caption{\it The average discounted cost of threshold combination $(\xi_1,\xi_2)$ after 2000 simulation paths.}
	\label{AS-8}
\end{figure}

There is still a way to determine the threshold of maintenance, that is to enumerate all possible threshold combinations. Based on the above model parameters, we examine the threshold combination about the degradation state of the system $(\xi_1,\xi_2)\in[0,5)\times[\xi_1,5]$. Clearly, $\xi_1\le\xi_2$. When $\xi_1=\xi_2$, we only consider the corrective maintenance action above the threshold. For computational convenience, we still take 0.1 as the interval, simulate the PDMP path 2000 times for all feasible threshold combinations, and calculate the average discounted cost. The result is shown in Figure \ref{AS-8}.

\begin{figure}[htbp]
	\centering
	\includegraphics[scale=0.5]{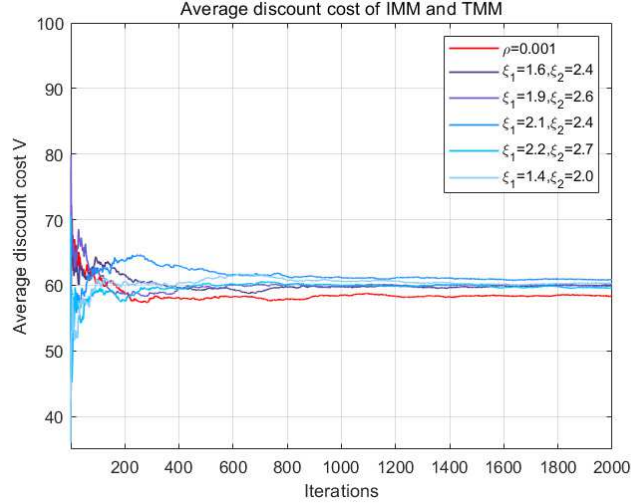}
	\caption{\it Average discounted cost of 2000 simulation paths of IMM and TMM with different threshold combinations.}
	\label{AS-9}
\end{figure}

It is obvious that the average discounted costs of different threshold combinations show the shape of a convex function, which indicates that there is a threshold combination to minimize the average discounted cost. In the above computation, the minimal threshold combination is $\xi_1 =2.2,\xi_2 = 2.7$ with average discounted cost equals to  59.57. Because there will be errors in calculating the average discounted cost by simulating 2000 paths, we choose five threshold combinations with the lowest average discount cost to compare with the optimal discounted cost calculated according to our model. The result is shown in Figure \ref{AS-9}.

The red line in Figure \ref{AS-9} represents the average discounted cost of IMM, the blue lines represent the average discounted  cost of TMM under different threshold combinations. After 2000 iterations, the average discounted cost of the IMM and TMM of the above five threshold combinations is given in Table\ref{table-1}. As shown in the table, the average discount cost of any threshold combination after 2000 iterations is higher than that of our model. This is due to the following reasons. First, the optimal action selection of our imperfect maintenance model is based on three variables, namely, the degradation state $w$, the number of executed imperfect maintenance $n$ and the current time $\theta$. Therefore, in order to obtain the optimal threshold strategy combination, the threshold combination of the above three variables $\{(\xi_{w1},\xi_{w2}),(\xi_{n1},\xi_{n2}),(\xi_{\theta1},\xi_{\theta2})\}$ should be selected for calculation. 
The previous simulation of threshold value only considered the combination of two thresholds about the damage degree $w$, and the minimum discounted cost is not globally optimal. Since there is no direct method to determine the optimal threshold combination, it is still necessary to enumerate all the possibilities mentioned above. When the ranges of $w,n,\theta$ are large, it would take a considerably long time to complete the computation. Second, by simulating 2000 PDMP paths and calculating the average discounted cost, only an approximation of the optimal discounted cost can be obtained; an exact optimal discounted cost cannot be guaranteed. When the number of iterations is small, such as 200 to 300, the average discounted cost fluctuates substantially, making it difficult to determine which threshold combination is the best. In order to get more accurate results, more iterations are needed. Third, after numerous simulations, the average discounted cost calculated in this way will still vary in a narrow range due to random shocks, a deterministic threshold is therefore impossible to obtain.

\begin{table}[htbp]
	\centering
	\title{}
	\caption{\it The average discounted cost of IMM and TMM with different threshold combinations.}
	\label{table-1}
	\setlength{\tabcolsep}{4mm}{
		\begin{tabular}{ccccccc}
			\toprule
			\multirow{2}{*}{\textbf{Model type}} & \multirow{2}{*}{IMM} & $\xi_1=1.6$      &  $\xi_1=1.9$     &  $\xi_1=2.1$     &  $\xi_1=2.2$     &  $\xi_1=1.4$     \\
			&  &  $\xi_2=2.4$ &  $\xi_2=2.6$     & $\xi_2=2.4$      &   $\xi_2=2.7$    &          $\xi_2=2.0$    \\\hline
			\textbf{$V$}                         & 58.06                & 60.05 & 59.90 & 60.83 & 59.57 & 60.29 \\
			\bottomrule
	\end{tabular}}
\end{table}

\begin{figure}[htbp]
	\centering
	\includegraphics[scale=0.4]{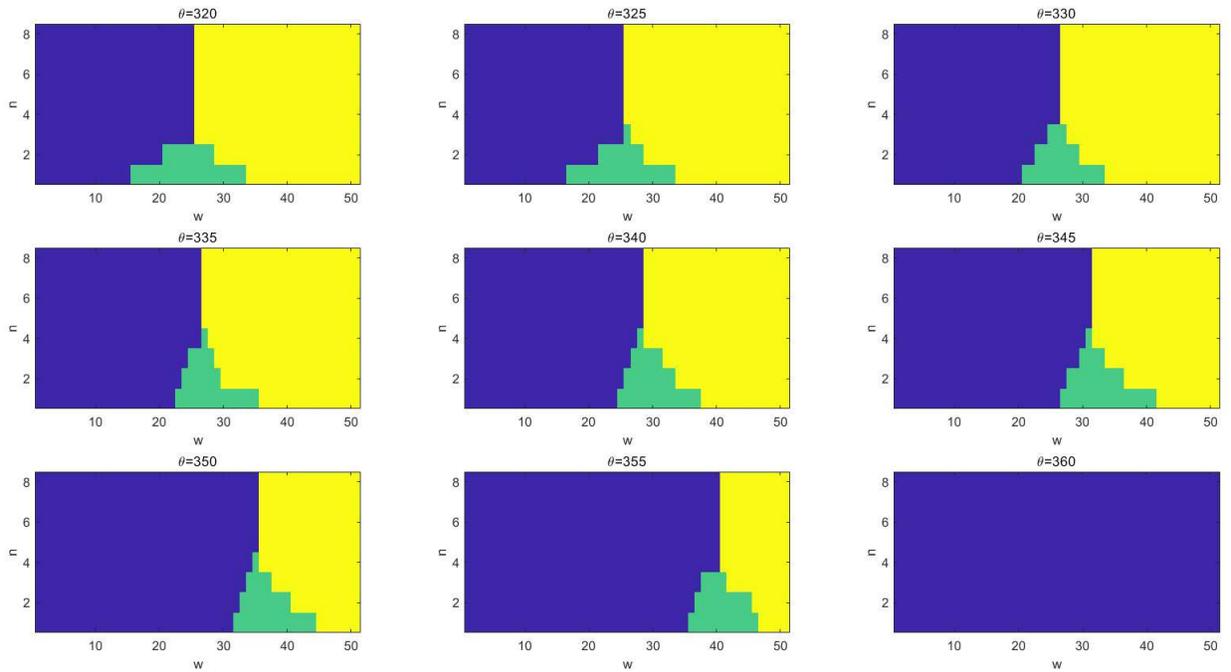}
	\caption{\it The changes of optimal actions when $\theta$ ranging from 320 to 360.}
	\label{AS-10}
\end{figure}

According to the definition of $\mathfrak{B}W$, it is easy to know that as the number of executed imperfect maintenances $n$ increases, the optimal strategy should tend to corrective maintenance; As time $\theta$ increases, the optimal strategy will tend to no maintenance. Figure \ref{AS-10} gives the optimal action distribution ranging from $\theta = 320$ to 360 with interval of 5 days. The blue, green, yellow colors represent no maintenance, imperfect maintenance and corrective maintenance respectively. Clearly, as time $\theta$ increases, the threshold of imperfect maintenance and corrective maintenance moves to the right. The threshold of $n$ increases first and then decreases. When $\theta = 320$, the dividing point of three actions is $w=2.6$, $n=2$. At this point, the optimal action is imperfect maintenance. When $\theta = 360$, the dividing point changes to $w=3.1$, $n=4$. And when $\theta = 355$, the dividing point becomes $w = 4.1$, $n=3$. This means the cost function is also convex with respect to $n$.

Our approach has the following advantages over TMM. First, our method can handle scenarios where it is challenging to express strategies using threshold models, such as circumstances where there are numerous different types of thresholds. Second, the computational accuracy of our model is better. The calculation of average discounted cost cannot avoid errors by exhausting all threshold combinations; hence, large-scale calculations are necessary to attain improved accuracy. As long as the iteration matrix converges by setting the stopping limit $\varepsilon$, our model can achieve the approximate optimal strategy. Third, the optimal strategy obtained by our model is deterministic while the optimal threshold combination obtained by exhausting all possible thresholds is unstable.

\section{Sensitivity analysis}\label{section5}
We perform a sensitivity analysis to determine how the optimal strategy will be affected by the uncertainty of our model's parameters. We discuss the influence of discount factor $\rho$, inspection time interval $T_{\mathrm{isp}}$ and maintenance costs $C_1,C_2$ on the optimal discounted cost and optimal strategy.

\subsection{Sensitivity analysis on discount factor}
\begin{figure}[htbp]
	\centering
	\includegraphics[scale=0.5]{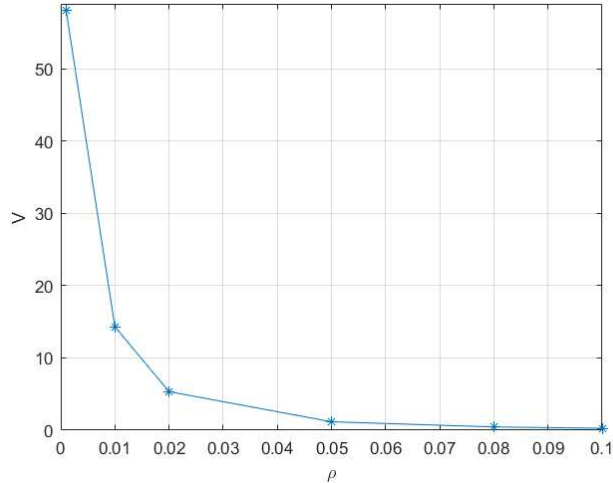}
	\caption{\it Change of discounted cost when the discount factor $\rho$ changing from 0.001 to 0.1. Here $T_{\mathrm{isp}} = 20$, $T_{\mathrm{soj}} = 5$, $T_{\mathrm{end}}=365$, $\alpha=1$, $\beta=1$, $\mu = 1/60$, $\lambda = (1/60)^2$, $C_1=\lfloor w \rfloor + n$, $C_2 =10$, $C_3=20$, $C_{\mathrm{isp}} = 1$.}
	\label{SA-1}
\end{figure}

Discount factor is crucial to the calculation of discounted cost. It displays the current discount rate for charges in the future. As the discount factor rises, the current value of the future cost falls; on the other hand, as the discount factor falls, the current value of the future cost rises closer to the value of the true future cost. There is no discount and the computed cost is equal to the total cost as the discount factor approaches zero.
Figure \ref{SA-1} shows the change of discounted cost when the discount factor changes from 0.001 to 0.1 with other parameters fixed. As shown in the figure, the discount cost decreases as $\rho$ raises. Table \ref{table-2} shows the values of discounted cost under different discount factors.

\begin{figure}[htbp]
	\centering
	\includegraphics[scale=0.5]{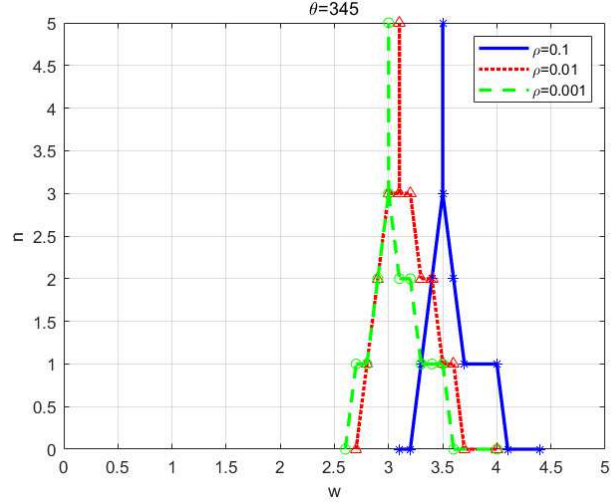}
	\caption{\it Change of the optimal action boundary when $\rho=0.1,0.01,0.001$ at $\theta = 345$.}
	\label{SA-2}
\end{figure}

As the discount factor changes, the optimal action will also change at the same time. When the discount factor is relatively small, the future cost will have a significant impact on the present cost. The decision-maker will be more likely to decide to perform maintenance now to keep the damage condition at a low level in order to avoid future costs of corrective maintenance.
Figure\ref{SA-2} gives the change of the optimal action boundary when $\rho=0.1,0.01,0.001$ at $\theta = 345$. Clearly, as $\rho$ becomes smaller, the optimal action boundary gradually moves to the left, which means that decision-maker is more inclined to plan maintenance actions.

\begin{table}[htbp]
	\centering
	\title{}
	\caption{\it Change of discounted cost when the discount factor $\rho$ changes from 0.001 to 0.1.}
	\label{table-2}
	\setlength{\tabcolsep}{4mm}{
		\begin{tabular}{ccccccc}
			\toprule
			$\rho$ & 0.001 & 0.01 & 0.02 & 0.05 & 0.08 & 0.1\\
			$V$ & 58.06 &14.32 & 5.34& 1.17& 0.45& 0.27\\
			\bottomrule
	\end{tabular}	}
\end{table}

\subsection{Sensitivity analysis on inspection time interval}
An essential factor taken into account in the majority of maintenance models is the ideal observation interval. The deterioration status of the system should not change much when the inspection time interval is too short. This would result in a lot of pointless inspections and raise the cost. The system is in a high state of deterioration when the inspection interval is too long, which raises operating costs and increases the likelihood of system failure. The foregoing analysis clearly shows that as the observation interval is increased, the system's overall cost first reduces before increasing. This implies that the cost is a convex function with regard to the observation interval. Therefore, an optimal interval could be chosen in order to keep the system's cost to a minimum.

\begin{figure}[htbp]
	\centering
	\includegraphics[scale=0.5]{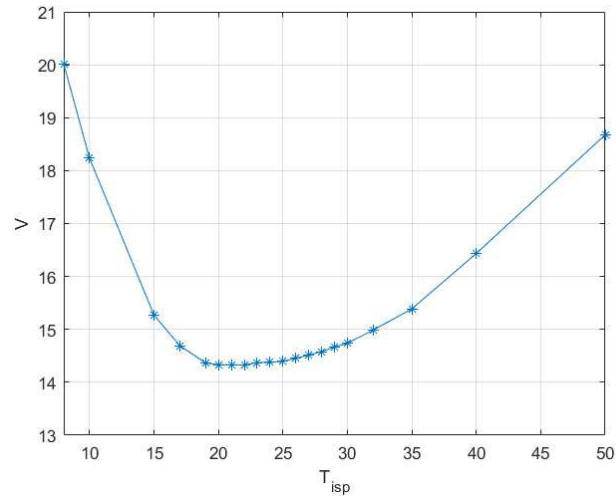}
	\caption{\it Change of average discounted cost when $T_{\mathrm{isp}}$ ranging from 8 to 50. Here $\rho = 0.01$, $T_{\mathrm{soj}} = 5$, $T_{\mathrm{end}}=365$, $\alpha=1$, $\beta=1$, $\mu = 1/60$, $\lambda = (1/60)^2$, $C_1=\lfloor w \rfloor + n$, $C_2 =10$, $C_3=20$, $C_{\mathrm{isp}} = 1$.}
	\label{SA-3}
\end{figure}

Based on the definition of $W_k$, as the discount factor $\rho$ increases, fewer iterations are necessary for a convergence to occur. 
For computational convenience, we choose $\rho = 0.01$, fix other parameters, and change the inspection interval $T_{\mathrm{isp}}$. We then calculate the corresponding optimal discounted cost, simulate the 2000 paths per interval, and compute the average discounted cost. The outcome is displaced in Figure \ref{SA-3}.

As shown in the figure, when $T_{\mathrm{isp}}\le 20$, the average discounted cost decreases quickly as $T_{\mathrm{isp}}$ increases. When $T_{\mathrm{isp}}$ ranges from 20 to 25, the average discounted cost does not change significantly. In our model, the average discounted cost is minimal when $T_{\mathrm{isp}}=21$. When $T_{\mathrm{isp}}\ge 25$, the average discounted cost increases gradually. This is in line with our previous judgment that the average discount cost is a convex function of inspection time $T_{\mathrm{isp}}$.

\begin{figure}[htbp]
	\centering
	\includegraphics[scale=0.5]{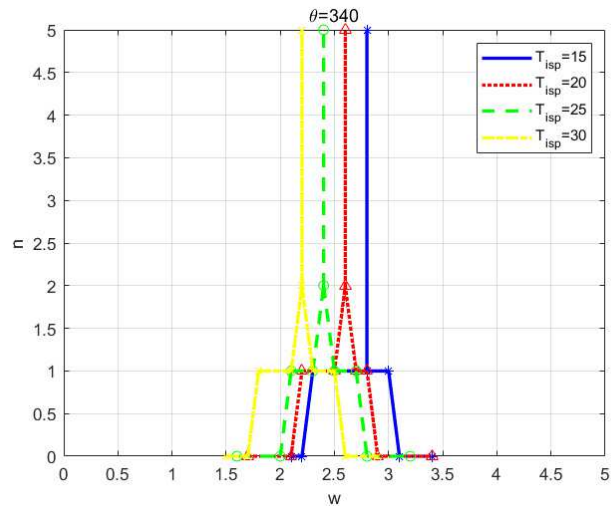}
	\caption{\it Change of optimal action boundary when $T_{\mathrm{isp}}=15,20,25,30$ at $\theta = 340$.}
	\label{SA-4}
\end{figure}

Figure \ref{SA-4} gives the change of optimal action boundary when $T_\mathrm{isp}=15,20,25,30$ at $\theta = 340$. It is evident that the optimal action border is progressively shifting to the left as the inspection interval is increased. This is also consistent with our regular judgment that since there are more observations, we may wait until the unit is in a worse condition of degradation before having it repaired. Due to natural deterioration and unpredictable shocks, the unit is more likely to enter the failure state the longer the observation interval. As a result, ``preventive maintenance" is increasingly essential, which simply means that even if the condition is generally in good shape, it is still required to plan the maintenance activity.

\subsection{Sensitivity analysis on maintenance costs}
In this subsection, we examine the impact of maintenance costs on discounted costs and optimal actions. We first investigate the situation of changing the fixed cost of imperfect maintenance, then study the situation of changing the corrective maintenance cost.

\begin{figure}[htbp]
	\centering
	\includegraphics[scale=0.5]{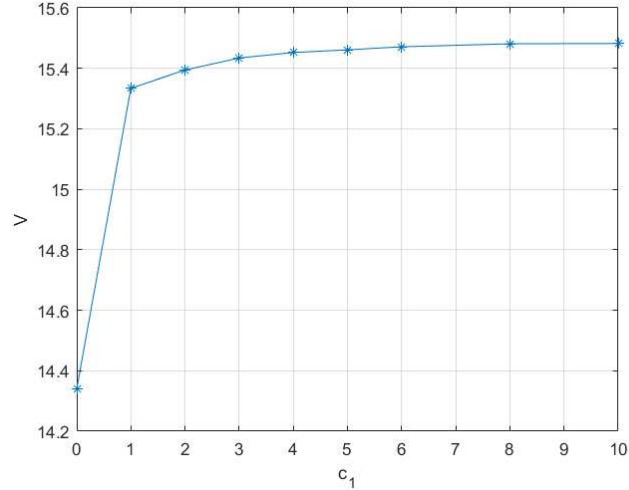}
	\caption{\it Change of average discounted cost when $c_1=0,1,2,3,4,5,8,10$ after 2000 simulation paths. Here $\rho = 0.01$, $T_{\mathrm{isp}} = 20$, $T_{\mathrm{soj}} = 5$, $T_{\mathrm{end}}=365$, $\alpha=1$, $\beta=1$, $\mu = 1/60$, $\lambda = (1/60)^2$, $C_1=c_1+\lfloor w \rfloor + n$, $C_2 =10$, $C_3=20$, $C_{\mathrm{isp}} = 1$.}
	\label{SA-5}
\end{figure}

\begin{figure}[htbp]
	\centering
	\includegraphics[scale=0.55]{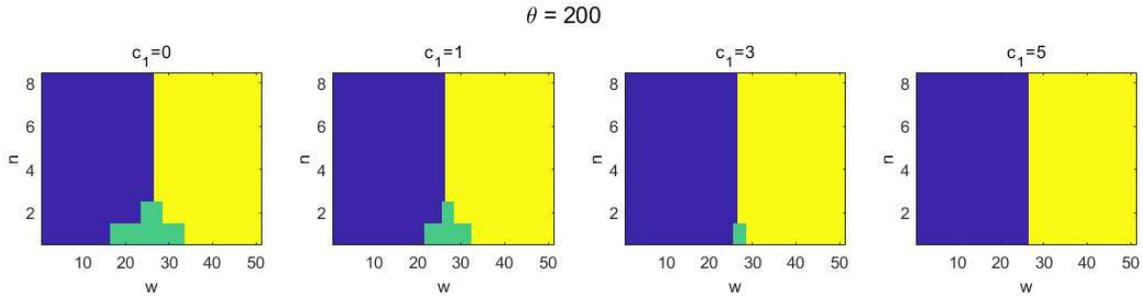}
	\caption{\it Optimal action distribution when $c_1=0,1,3,5$ at $\theta = 200$. }
	\label{SA-6}
\end{figure}

The area of imperfect maintenance action is significantly affected by the fixed imperfect maintenance cost. The action area of imperfect maintenance will be reduced in proportion to the increase of the cost.
We define the imperfect maintenance cost as $C_1 = c_1 + \lfloor w \rfloor + n$, where $c_1$ is the fixed cost. In the previous simulation, the fixed cost is assumed to be zero. We now vary the fixed cost $c_1$ and remain other parameters unchanged. The simulation result is given in Figure \ref{SA-5}.

As shown in the figure, when the fixed cost $c_1$ increases, the average discounted cost gradually rises and finally becomes stable. In fact, with the increase of fixed cost $c_1$, the cost of imperfect maintenance is rising. 
The attraction of imperfect maintenance is progressively fading in comparison to corrective maintenance. The drawback of imperfect maintenance, that it cannot restore the unit to an ideal state, has steadily grown more pronounced. The decision-maker will progressively go toward planning corrective maintenance because the use of imperfect maintenance will increase the number of maintenance. The decision-maker will only consider no maintenance and corrective maintenance when the anticipated cost of imperfect maintenance is higher than that of corrective maintenance.
The illustration of distributions of optimal actions when $c_1=0,1,3,5$ at $\theta = 200$ is given in Figure \ref{SA-6}. It can be clearly seen that, as $c_1$ increases, the area of imperfect maintenance is decreasing. When $c_1=5$, The optimal options left are no maintenance and corrective maintenance. 

\begin{figure}[htbp]
	\centering
	\includegraphics[scale=0.5]{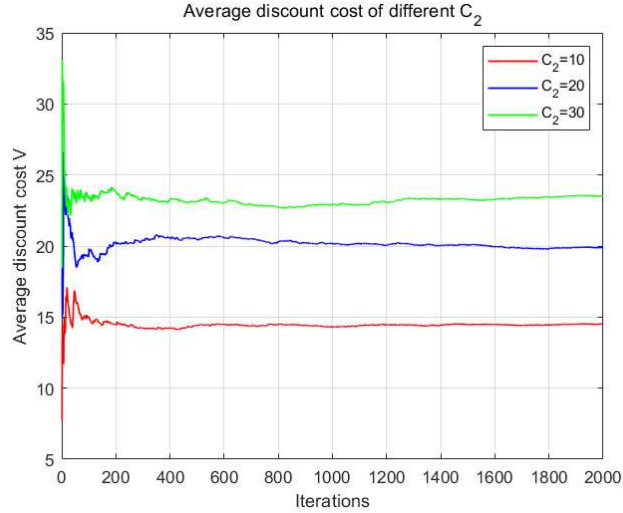}
	\caption{\it Average discounted cost of $C_2=10,20,30$ after 2000 simulation paths. Here $\rho = 0.01$, $T_{\mathrm{isp}} = 20$, $T_{\mathrm{soj}} = 5$, $T_{\mathrm{end}}=365$, $\alpha=1$, $\beta=1$, $\mu = 1/60$, $\lambda = (1/60)^2$, $C_1=\lfloor w \rfloor + n$,$C_3=40$, $C_{\mathrm{isp}} = 1$.}
	\label{SA-7}
\end{figure}

\begin{figure}[htbp]
	\centering
	\includegraphics[scale=0.5]{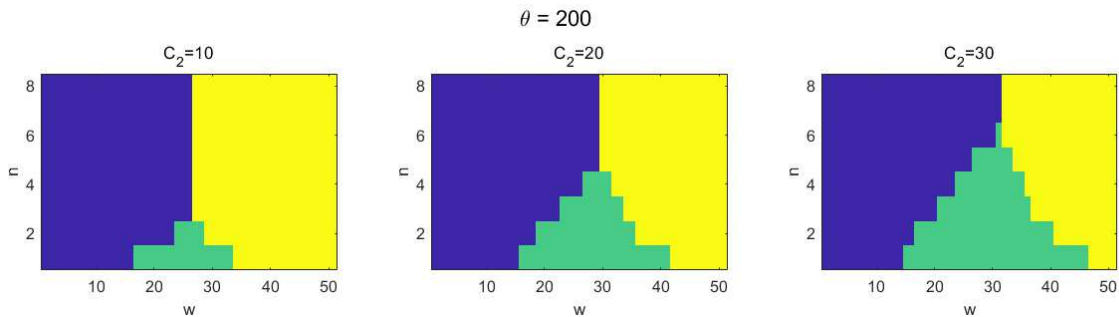}
	\caption{\it Optimal action distribution when $C_2=10,20,30$ at $\theta = 200$.}
	\label{SA-8}
\end{figure}

In contrast to the fixed cost of imperfect maintenance, the decision-maker will be more likely to select imperfect maintenance when the cost of corrective maintenance rises.
Since $C_3$ is always greater than $C_2$, in order to better reflect the impact of changing maintenance costs, we let $C_1 = \lfloor w\rfloor + n$, $C_3 = 40$,  change $C_2 = 10,20,30$, and fix other parameters unchanged. The result is shown in Figure \ref{SA-7}. 
In the picture, the discounted cost is gradually rising along with the expense of corrective maintenance. When $C_2$ is high, the discounted cost will typically be the best discounted cost that simply uses no maintenance or imperfect maintenance as a strategy.

The optimal action allocation for various corrective maintenance costs is shown in Figure \ref{SA-8}. It is evident that when the cost of corrective maintenance rises, the imperfect maintenance action area is gradually growing, while the corrective maintenance action area and the no maintenance action area are gradually contracting. As the cost of corrective maintenance increases, the system will incur significant costs as long as it is in the failure state even once because it could only adopt corrective maintenance while it is in the failure state. In this scenario, decision-makers will be more prone to trying to prolong the system's reduced deterioration condition. As a result, the imperfect maintenance action area will grow while the no maintenance action area will decrease.

\section{Conclusion}\label{section6}
In this article, a condition-based imperfect maintenance model is constructed using PDMP. The system deteriorates as a result of random shocks and natural degradation. The natural degradation is deterministic can be described by ODEs. The damage increment from a random shock follows an IG distribution. The parameters of IG distribution are related to the jump intensity of random shocks. The inspection time is fixed. At the time of inspection, the decision-maker will schedule a maintenance task according to the current state of the system. 
The system continues to deteriorate while waiting for maintenance. Maintenance strategies include corrective maintenance and imperfect maintenance. For imperfect maintenance, the improvement factor model is adopted, in which the improvement factor follows a Beta distribution and the parameters of the Beta distribution are correlated with the total number of performed imperfect maintenances. 
In order to minimize the total discounted cost, the decision-maker needs to make a choice among no maintenance, imperfect maintenance and perfect maintenance at each inspection time.  

Due to the complexity of the model, the C-K equation derived is incredibly difficult to solve. The optimal impulse control theory in \cite{Costa1} is applied by creating iterative equations (\ref{thm1-1}) to solve the model. An illustrative study of a component coating maintenance problem is presented to further explain the model.

The threshold technique is frequently used to identify the best maintenance procedures. It is impossible to directly determine the ideal thresholds given the complexity of our model. There are three advantages of our method over the one acquired by repeatedly iterating through all possible threshold combinations. First of all, we can address issues where it is challenging to express threshold values. Second, it facilitates calculation when there are multiple threshold values. Third, the chosen of optimal actions is deterministic.

There are still certain issues that require further research. For instance, one can discuss, how would the optimal action change when the damage increment subject to different distributions.
Additionally, since the observation in our model is thorough and sufficient, partially observable PDMP models are worthwhile taking into account.
Meanwhile, models with long-term average cost can be considered because only discounted cost are discussed in our work.
Of course, the building and analysis of the relevant model would be more challenging. Further novel and effective methods are still worth exploring and developing.

\section*{CRediT authorship contribution statement}
{\bf Weikai Wang:} Conceptualization, Methodology, Investigation, Software, Validation, Formal analysis, Writing - original draft. {\bf Xian Chen:} Conceptualization, Methodology, Validation, Formal analysis, Writing - review and editing, Supervision. 

\section*{Declaration of competing interest}
The authors declare that they have no known competing financial interests or personal relationships that could have appeared to influence the work reported in this paper.


\begin{thebibliography}{99}
	\bibitem{Alaswad1} S. Alaswad and Y. Xiang, ``A review on condition-based maintenance optimization models
	for stochastically deteriorating system,” \textit{Reliability Engineering and System Safety}, vol. 157,
	pp. 54–63, 2017.
	
	\bibitem{Pham1} H. Pham and H. Wang, ``Imperfect maintenance,” \textit{European Journal of Operational Research},
	vol. 94, no. 3, pp. 425–438, 1996.
	
	\bibitem{Huynh1} K. Huynh, I. Castro, A. Barros, et al., ``Modeling age-based maintenance strategies with minimal
	repairs for systems subject to competing failure modes due to degradation and shocks,”
	\textit{European Journal of Operational Research}, vol. 218, no. 1, pp. 140–151, 2012.
	
	\bibitem{WuF1} F. Wu, S. Niknam, and J. Kobza, ``A cost effective degradation-based maintenance strategy
	under imperfect repair,” \textit{Reliability Engineering and System Safety}, vol. 144, pp. 234–243,
	2015.
	
	\bibitem{Mosayebi1} E. Mosayebi Omshi and A. Grall, ``Replacement and imperfect repair of deteriorating system:
	Study of a CBM policy and impact of repair efficiency,” \textit{Reliability Engineering and System
	Safety}, vol. 215:107905, 2021.
	
	\bibitem{LiYanfu1} J. Xu, Z. Liang, Y. Li, et al., ``Generalized condition-based maintenance optimization for
	multi-component systems considering stochastic dependency and imperfect maintenance,”
	\textit{Reliability Engineering and System Safety}, vol. 211:107592, 2021.
	
	\bibitem{Malik1} M. Malik, ``Reliable preventive maintenance scheduling,” \textit{AIIE Transactions}, vol. 11, no. 3,
	pp. 221–228, 1979.
	
	\bibitem{DoP1} P. Do, A. Voisin, E. Levrat, et al., ``A proactive condition-based maintenance strategy with
	both perfect and imperfect maintenance actions,” \textit{Reliability Engineering and System Safety},
	vol. 133, pp. 22–32, 2015.
	
	\bibitem{WangJH1} J. Wang, X. Zhang, J. Zeng, et al., ``Optimal dynamic imperfect preventive maintenance of
	wind turbines based on general renewal processes,” \textit{Internal Journal of Production Research},
	vol. 58, no. 22, pp. 6791–6810, 2020.
	
	\bibitem{Davis1} M. Davis, ``Piecewise-deterministic Markov processes: A general class of non-diffusion stochastic models,” \textit{Journal of the Royal Statistical Society: Series B, Statistical Methodology}, vol. 46, no. 3, pp. 353–376, 1984.
	
	\bibitem{ChenXian1} X. Chen and C. Jia, ``Limit theorems for generalized density-dependent Markov chains and
	bursty stochastic gene regulatory networks,” \textit{Journal of Mathematical Biology}, vol. 80, no. 4,
	pp. 959–994, 2019.
	
	\bibitem{Bauerle1} N. B\"{a}uerle and U. Rieder, ``MDP algorithms for portfolio optimization problems in pure jump
	markets,” \textit{Finance and Stochastics}, vol. 13, no. 4, pp. 591–611, 2009.
	
	\bibitem{Defourny1} S. Moazenia and B. Defourny, ``Optimal control of energy storage under random operation
	permissions,” \textit{IISE Transactions}, vol. 50, no. 8, pp. 2472–5862, 2018.
	
	\bibitem{Dufour1} C. Pasin, F. Dufour, L. Villain, et al., ``Controlling IL-7 injections in HIV-infected patients,” \textit{Bulletin of Mathematical Biology}, vol. 80, pp. 2349–2377, 2018.
	
	\bibitem{Boukas1} E. Boukas, Q. Zhu, and Q. Zhang, ``Piecewise deterministic markov process model for flexible
	manufacturing systems with preventive maintenance,” \textit{Journal of Optimization Theory and
	Applications}, vol. 81, no. 2, pp. 259–275, 1994.
	
	\bibitem{Dufour2} H. Zhang, F. Innal, F. Dufour, et al., ``Piecewise deterministic Markov processes based approach applied to an offshore oil production system,” \textit{Reliability Engineering and System
	Safety}, vol. 126, pp. 126–134, 2014.
	
	\bibitem{Lair1} W. Lair, S. Mercier, M. Roussignol, et al., ``Piecewise deterministic Markov processes and
	maintenance modelling: application to maintenance of a train air-conditioning system,” \textit{Proceedings
	of the Institution of Mechanical Engineers, Part O: Journal of Risk and Reliability}, vol. 225, pp. 199–209, 2011.
	
	\bibitem{Demgne1} J. Demgne, S. Mercier, W. Lair, et al., ``Modelling and numerical assessment of a maintenance
	strategy with stock through piecewise deterministic Markov processes and quasi Monte Carlo
	methods,” \textit{Proceedings of the Institution of Mechanical Engineers, Part O: Journal of Risk
	and Reliability}, vol. 231, no. 4, pp. 429–445, 2017.
	
	\bibitem{LiYanfu2} Y. Lin, Y. Li, and E. Zio, ``A framework for modeling and optimizing maintenance in systems
	considering epistemic uncertainty and degradation dependence based on PDMPs,” \textit{IEEE
	Transactions on Industrial Informatics}, vol. 14, no. 1, pp. 210–220, 2018.
	
	\bibitem{Arismendi1}R. Arismendi, A. Barros, and A. Grall, ``Piecewise deterministic Markov process for condition-based maintenance models—Application to critical infrastructures with discrete-state deterioration,”
	\textit{Reliability Engineering and System Safety}, vol. 212:107540, 2021.
	
	\bibitem{Fatima1} T. Fatima and A. Muntean, ``Sulfate attack in sewer pipes: Derivation of a concrete corrosion
	model via two-scale convergence,” \textit{Nonlinear Analysis: Real World Applications}, vol. 15,
	pp. 326–344, 2014.
		
	\bibitem{Arya1}R. Arya, ``Finite element solution of coupled-partial differential and ordinary equations in
	multicomponent polymeric coatings,” \textit{Computers and Chemical Engineering}, vol. 50, pp. 152–
	183, 2013.
	
	\bibitem{Karanci1}E. Karanci and R. Betti, ``Modeling corrosion in suspension bridge main cables. II: Long-term
	corrosion and remaining strength,” \textit{Journal of Bridge Engineering}, vol. 23, no. 6, pp. 1–15,
	2018.
	
	\bibitem{Panchenko1} Y. Panchenko and A. Marshakov, ``Long-term prediction of metal corrosion losses in atmosphere using a power-linear function,” \textit{Corrosion Science}, vol. 109, pp. 217–229, 2016.
	
	\bibitem{Panchenko2}Y. Panchenko, A. Marshakov, L. Nikolaeva, et al., ``Long-term prediction of corrosion losses of metals by means of various functions for the continental part of Russia,” \textit{Protection of Metals
	and Physical Chemistry of Surfaces}, vol. 54, no. 7, pp. 1266–1275, 2018.

	\bibitem{Costa1} O. Costa, F. Dufour, and A. Piunovskiy, ``Constrained and unconstrained optimal discounted
	control of piecewise deterministic Markov processes,” \textit{SIAM Journal of Control and Optimization},
	vol. 54, no. 3, pp. 1444–1474, 2016.
	
	\bibitem{YeZS1}Z. Ye and N. Chen, ``The inverse Gaussian process as a degradation model,” \textit{Technometrics},
	vol. 56, pp. 302–311, 2013.
	
	\bibitem{WuD1}D. Wu, R. Peng, and S. Wu, ``A review of the extensions of the geometric process, applications,
	and challenges,” \textit{Quality and Reliability Engineering International}, vol. 36, no. 2, pp. 436–446,
	2020.
	
	\bibitem{Costa2} O. Costa and M. Davis, ``Impulse control of piecewise deterministic processes,” \textit{Mathematics of Control, Signals, and Systems}, vol. 2, pp. 187–206, 1989.
	
	\bibitem{Dufour3} B. Saporta and F. Dufour, ``Numerical method for impulse control of piecewise deterministic
	Markov processes,” \textit{Automatica}, vol. 48, pp. 779–793, 2012.
	
	\bibitem{Dufour4} B. Saporta, F. Dufour, and A. Geeraert, ``Optimal strategies for impulse control of piecewise
	deterministic Markov processes,” \textit{Automatica}, vol. 77, pp. 219–229, 2017.
	
	\bibitem{LiYX1} Y. Li, Y. Zhang, S. Jungwirth, et al., ``Corrosion inhibitors for metals in maintenance equipment: introduction and recent developments,” \textit{Corrosion Reviews}, vol. 32, no. 5-6, pp. 163–
	181, 2014.

\end{thebibliography}
\end{document}